\documentclass{article}

\usepackage{arxiv}

\usepackage[utf8]{inputenc} 
\usepackage[T1]{fontenc}    
\usepackage{hyperref}       
\usepackage{url}            
\usepackage{booktabs}       
\usepackage{amsfonts}       
\usepackage{nicefrac}       
\usepackage{microtype}      
\usepackage{graphicx}

\usepackage{amsmath}
\usepackage{multirow}
\usepackage{graphicx}
\usepackage{subcaption}
\usepackage{float}
\usepackage{authblk}
\usepackage[%
    style=numeric-comp,sorting=none,
    sortcites=true,doi=false,url=false,
    giveninits=true,hyperref]{biblatex}
\makeatletter
\g@addto@macro{\UrlBreaks}{\UrlOrds}
\makeatother

\DeclareUnicodeCharacter{2215}{/}

\title{Anisotropy and Current Control of Magnetization in SrRuO$_3$/SrTiO$_3$ Heterostructures for Spin-Memristors}

\author[1,2]{ Anouk S.~Goossens }
\author[1,2]{ M. A. T. Leivisk{\"a} }
\author[1,2]{ T.~Banerjee }
\affil[1]{Groningen Cognitive Systems and Materials Center, University of Groningen, Groningen, The Netherlands}
\affil[2]{Zernike Institute for Advanced Materials, University of Groningen, Groningen, The Netherlands}
\affil[ ]{\texttt{\{a.s.goossens,t.banerjee\}@rug.nl}}

\begin{document}
\maketitle

\begin{abstract}
Spintronics-based nonvolatile components in neuromorphic circuits offer the possibility of realizing novel functionalities at low power. Current-controlled electrical switching of magnetization is actively researched in this context. Complex oxide heterostructures with perpendicular magnetic anisotropy (PMA), consisting of SrRuO$_3$ (SRO) grown on SrTiO$_3$ (STO) are strong material contenders. Utilizing the crystal orientation, magnetic anisotropy in such simple heterostructures can be tuned to either exhibit a perfect or slightly tilted PMA. Here, we investigate current induced magnetization modulation in such tailored ferromagnetic layers with a material with strong spin-orbit coupling (Pt), exploiting the spin Hall effect. We find significant differences in the magnetic anisotropy between the SRO/STO heterostructures, as manifested in the first and second harmonic magnetoresistance measurements. Current-induced magnetization switching can be realized with spin-orbit torques, but for systems with perfect PMA this switching is probabilistic as a result of the high symmetry. Slight tilting of the PMA can break this symmetry and allow the realization of deterministic switching. Control over the magnetic anisotropy of our heterostructures therefore provides control over the manner of switching. Based on our findings, we propose a three-terminal spintronic memristor, with a magnetic tunnel junction design, that shows several resistive states controlled by electric charge. Non-volatile states can be written through SOT by applying an in-plane current, and read out as a tunnel current by applying a small out-of-plane current. Depending on the anisotropy of the SRO layer, the writing mechanism is either deterministic or probabilistic allowing for different functionalities to emerge. We envisage that the probabilistic MTJs could be used as synapses while the deterministic devices can emulate neurons.
\end{abstract}


\section{Introduction}
Current information processing technologies make use of the von Neumann architecture. While this approach is suitable for performing simple iterative operations, it is not as well equipped for more complex tasks such as recognition and prediction. The brain, however, can tackle these tasks with considerable ease and for this reason, progressively more research is being conducted into brain-like computation. The field of artificial intelligence, for example, has seen great success in developing software that emulate the functioning of brain. However, these software still run on von Neumann hardware, making this approach the subject of several limitations, such as high power usage. Development of novel hardware that can directly mimic brain-like neural networks is therefore of significant interest. The main components of the brain are neurons, which process information, and synapses, which learn and remember by adjusting their connection strength. In particular, neurons are often regarded as stochastic spiking elements that transmit digital signals across analog synapses, whose conductance can take on a range of values (\cite{hwang2018,kwon2020}).

Spintronics, where information can be stored in a non-volatile way using magnetic states, offers an attractive platform for emulating these brain components in solid-state devices. Here, we focus on systems with perpendicular magnetic anisotropy (PMA), where the magnetization can be either parallel or antiparallel (pointing 'up' or 'down') to the film normal, making them suitable candidates for binary memory elements. PMA systems have various advantages over their in-plane easy axis counterparts, including enhanced the thermal stability of the magnetic ordering, higher data storage density, improved scalability and more facile writing of the states \cite{iwasaki1977,Tudu2017}.
Realizing analog states with the aid of magnetic domains \cite{Fukami2018} is also possible. The presence of the domains allows obtaining magnetic states where some domains point up and some point down so that the overall magnetization assumes a magnitude lower than the saturation value. The possibility of achieving analog states, largely absent in standard electronics, is particularly useful for emulating brain-like behavior which is believed to make use of a combination of analog and digital signals (\cite{shu2006}. 

In thin films, magnetization typically prefers to lie in the plane of the film due to the demagnetising energy cost associated with the out-of-plane orientation. Realizing a system with PMA therefore requires a compensation of this energy cost. In ferromagnetic/heavy metal multilayers, for example, the translational invariance lost in the direction of the film normal gives rise to an additional interfacial anisotropy. This can overcome the demagnetizing energy and yields PMA in the ferromagnet. However, this is an interfacial effect and hence limited to ultrathin ferromagnetic films, which can be challenging to fabricate \cite{lairson1994application,donzelli2003perpendicular,bandiera2012enhancement}. Crystalline alloys of Fe and Co, such as FePt, FePd and CoPt have also been investigated as PMA materials. In these materials the L1$_0$ phase has been found to have a large uniaxial magnetic anisotropy resulting in an out-of-plane magnetization \cite{zhang2010l10,seki2003,mizukami2011fast}. A drawback of these materials, however, is that their magnetization dynamics suffer from strong damping and hence significant energy dissipation. This aspect renders them less suitable for technological applications \cite{mizukami2011fast,iihama2014low,Tudu2017}. Other material systems exhibiting PMA include the ferromagnetic Heusler alloys \cite{wu2009epitaxial,mizukami2011long} with uniaxial anisotropy favoring out-of-plane magnetization, and amorphous rare-earth transition metal alloys \cite{cheng1989,okamine1985,harris1992}, where the combination of strong exchange interactions and strong magnetic anisotropy allows tailoring of the latter.

In order to write the magnetic state, it is important to be able to switch the magnetization orientation. When the ferromagnet is placed adjacent to a heavy metal layer, its magnetization orientation can be switched with an in-plane current owing to the presence of spin-orbit torques induced by the spin Hall effect and Rashba fields \cite{Miron2011,Liu2012,Liu2012b}. SOT switching has also been realized in a single-crystalline ferromagnetic layer with intrinsic bulk inversion asymmetry, such as GaMnAs \cite{jiang2019}. Here the Dresselhaus- and Rashba-like fields give rise to current-induced torque. A potential downside of the latter for memristive applications is that the single crystalline nature of the ferromagnet does not allow for intermediate magnetization states (between up and down) to be realized.

Systems with perfect PMA are symmetric with respect to these torques, rendering the magnetization switching probabilistic. Deterministic switching, on the other hand, can be realized by breaking the symmetry in the system. The simplest way of breaking this symmetry is via the application of a small in-plane field, but this is not an ideal solution from the viewpoint of technological applications. Other approaches include introducing asymmetry through exchange bias \cite{VanDenBrink2016,Fukami2016}, device geometries with lateral structural asymmetry inducing an anisotropy gradient \cite{Yu2014}, or through a precise control of the magnetocrystalline anisotropy of the ferromagnet \cite{Eason2013,You2015,Liu2019}.

Transition metal complex oxides have the unique potential for controlling their magnetocrystalline anisotropy via relatively straightforward means due to the intimate relationship between their crystal structure and magnetic properties. In this work we illustrate this potential with SrRuO$_3$ (SRO). SRO is a ferromagnetic metal with a relatively high Curie temperature around 160 K in bulk and between 130-150 K in thin films. It has a perovskite crystal structure accompanied by a slight rotation of the oxygen octahedron, which yields an orthorhombic unit cell with lattice constants of a = 5.5670 \AA, b = 5.5304 \AA, c = 7.8446 \AA. However, with the rotation being very small, the crystal structure can also be described as pseudocubic (pc) with a lattice constant of 3.93 \AA. Fig. S1 shows the relative orientations between the octahedral and pseudocubic unit cell notations. Throughout this text, the pseudocubic notation is used, indicated by a subscript $pc$. 

SRO is well-known for its strong magnetocrystalline anisotropy (anisotropy field $\sim$10 T) (\cite{Koster2012a}) stemming from the strong spin-orbit coupling of the heavy Ru ions. The magnetic easy axis of SRO has been observed to lie close to the crystallographic [110]$_{pc}$ ([010]$_o$) direction \cite{Marshall1999}. For this reason, the orientation of the unit cell will effectively determine the easy axis direction. Epitaxial thin films of SRO can be grown on SrTiO$_3$ (STO) substrates, as the latter has a perovskite structure with a closely matching lattice constant. Interestingly, the SRO unit cell orientation tends to adopt that of the STO substrate to minimize the in-plane strain \cite{jiang1998microstructure}. This provides us with an elegant tool for controlling the magnetocrystalline anisotropy of the SRO film without the requirement of ultrathin films, additional layers, or complex layer structures and device geometries. 

Here, we have utilized STO substrates with two different orientations, (001) and (110), where the Miller indices refer to the crystal planes parallel to the surface. As discussed above, the SRO films grown on these substrates are expected to adopt (001) and (110) orientations, respectively. Referring to the relationship between the crystal structure and the magnetocrystalline anisotropy, we would anticipate that the magnetic easy axis to lie along the film normal for the (110) film, while in the (001) film it is predicted to have an in-plane component. Due to the absence (presence) of this in-plane component, we expect the former (latter) to exhibit probabilistic (deterministic) switching. In order to enable current-induced switching of the SRO layer, a layer of strongly spin-orbit coupled Pt has been included in our device structures to serve as a source of spin-orbit torque. The heterostructures with different magnetocrystalline anisotropies and switching properties investigated in this work can potentially serve as spintronic memristive materials. By incorporating them in a three-terminal magnetic tunnel junction design, deterministic switching devices can emulate a spin-synapse whereas probabilistic switching devices can function as neurons. 
\section{Materials and Methods}

STO substrates with (001) and (110) orientations with small miscut angles were treated with buffered hydrofluoric acid and annealed in oxygen. We deposited SRO films on the STO substrates by pulsed laser deposition (PLD). A KrF excimer laser beam ($\lambda$ = 248 nm) was focused onto an SRO target at a repetition rate of 1 Hz, with a laser fluence of 1.5 Jcm$^{-2}$. The films were deposited at 600 $^{\circ}$C with an oxygen pressure of 0.13 mbar. After 1200 laser pulses, we cooled the films down to room temperature under an oxygen pressure of 100 mbar \cite{zhang2020}. Throughout this paper, we will refer to the SRO/STO (001) samples as SRO$_{001}$ and to the SRO/STO (110) samples as SRO$_{110}$.

We performed structural and magnetic characterization measurements on the films before fabricating devices. To determine the unit cell orientation of the SRO films on the different substrates we used different x-ray diffraction (XRD) techniques at room temperature. X-ray reflectivity measurements were used to estimate the thicknesses of the SRO layers. 2$\theta$-$\theta$ scans were performed to determine the out-of-plane lattice constant. We also conducted reciprocal space mapping (RSM) to infer information about the in-plane lattice parameters of the SRO films. Finally, azimuthal scans, where a diffraction peak appears whenever a reciprocal lattice point is crossed, were carried out and information about the film alignment relative to that of the substrate was inferred from their relative peak positions. For identical alignments, identical peak positions are expected. 

Temperature and field dependent magnetic measurements were performed using a superconducting quantum interference device (SQUID) magnetometer (Quantum design, MPMS). The different field orientations adopted were 0$^{\circ}$ (in the plane of the film), 90$^{\circ}$ (along the film normal) or at $\sim$45$^{\circ}$ (diagonal to the film normal).

Post structural and magnetic characterization, we sputtered a Pt layer on the SRO films for current-induced switching studies. The Pt/SRO bilayers were etched into Hall bars oriented along different crystallographic axes using UV lithography and ion beam etching techniques. Schematic diagrams of the Hall bars are shown in Fig. \ref{fig2} and Fig. \ref{fig3}, where the SRO and Pt layers are indicated in red and blue respectively. Electrical measurements were done by applying a alternating current (ac) along the x axis of the Hall bar. The first and second harmonic voltage responses were measured simultaneously in both the longitudinal (along the x axis parallel to the current) and transverse (along the y axis perpendicular to the current) direction. It should be noted that xz and yz scans were conducted on Hall bars of different, perpendicular orientation.

\section{Results}
\begin{figure}
    \centering
    \includegraphics[width=\textwidth]{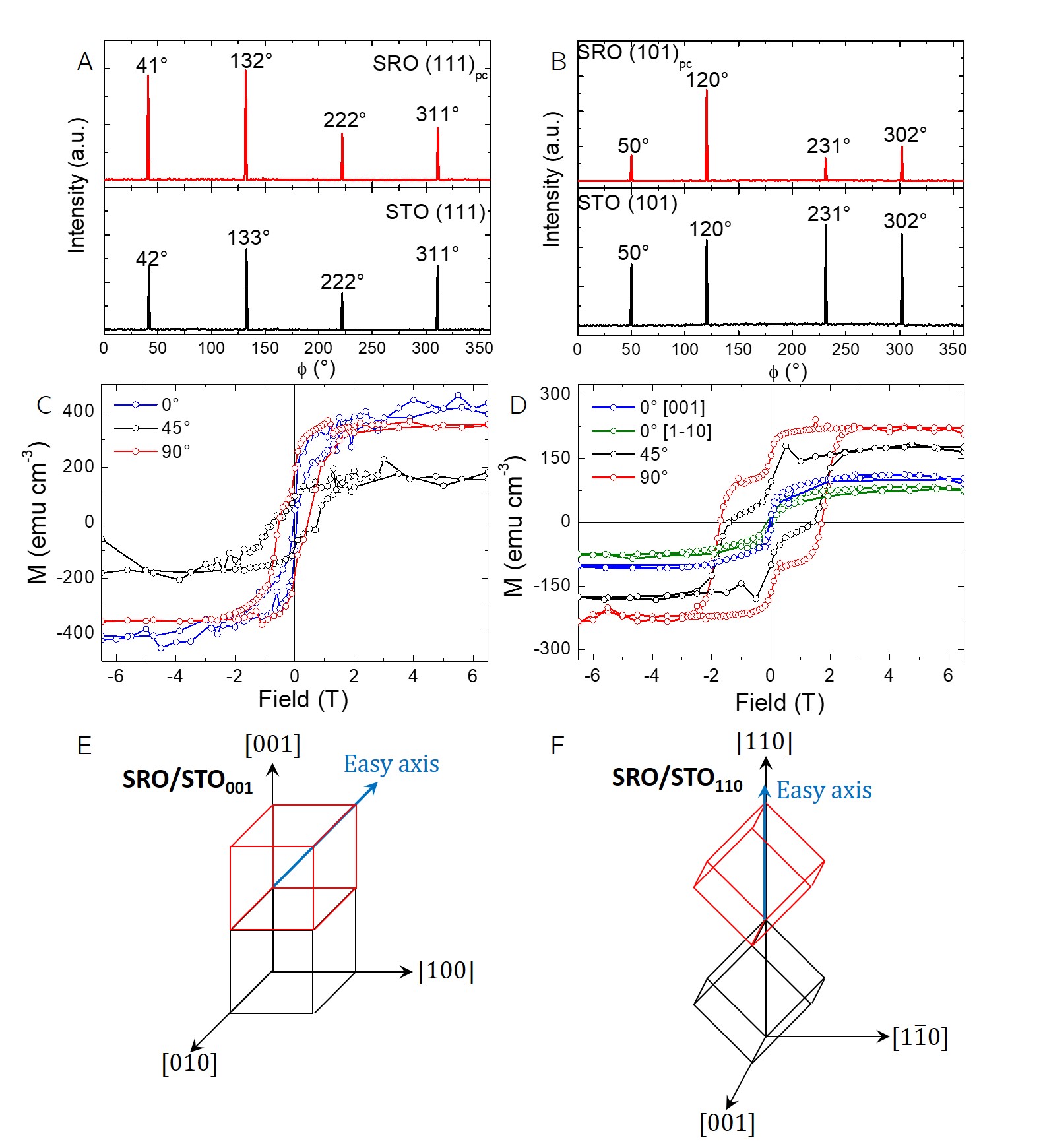}
    \caption{Azimuthal scans \textbf{(A)} for SRO/STO (001) around the SRO (111)$_{pc}$ and STO (111) peaks \textbf{(B)} for SRO$_{110}$ around the SRO (101)$_{pc}$ and STO (101) peaks. Magnetic field versus magnetization measurements at 10 K taken with the field at different angles with respect to in-plane for \textbf{(C)} SRO$_{001}$ and \textbf{(D)} SRO$_{110}$. \textbf{(E)} and \textbf{(F)} summarize the relative alignment between the unit cell of the substrate and the pseudocubic unit cell of the SRO film for SRO$_{001}$ and SRO$_{110}$ respectively. The figures also indicate the approximate direction of the magnetic easy axis.}
    \label{fig1}
\end{figure}
The 2$\theta$-$\theta$ scans (Fig. S4A) confirm the (001) orientations of both the STO substrate and SRO film. We calculated the SRO out-of-plane lattice constant to be 3.968 \AA, revealing a tensile strain of 0.97$\%$. Similarly for the sample grown on STO (110) substrate (Fig S4B), the 2$\theta$-$\theta$ diffraction peaks correspond to the {110} crystal planes, suggesting that both STO and SRO have (110) orientation. For SRO, the spacing between (110) planes was found to be 2.788 \AA and slightly deviates from the bulk value of 2.779 \AA, suggesting a small tensile strain.

The Kiessig fringes of SRO$_{001}$ (Fig. S3A) at low angles could be fit using the recursive Parratt formalism giving a thickness estimate of 10.7 nm. For SRO$_{110}$ (Fig. S3B) no good fit could be obtained, which is consistent with the larger surface roughness of 2 nm for SRO$_{110}$ compared to the 0.1 nm roughness of the SRO$_{001}$ surface (as seen in AFM images shown in Fig. S2). The thickness of SRO$_{110}$ was estimated to be close to 11 nm by measuring the step height between the film and substrate with AFM. 

The 3D construction of the reciprocal space maps (Fig. S4 C and D) for both samples shows a vertical alignment of the STO and SRO peaks, indicating that the films are fully strained in-plane to match the substrate lattice constant. As a result, both films experience an in-plane compressive strain of 0.64$\%$.

For SRO$_{001}$, azimuthal scans were conducted around the (001), (101) (Fig. \ref{fig1}A), and (111) lattice planes. In most cases, the film (top panel) and substrate (bottom panel) peaks are seen to closely coincide. The only discrepancy is for the (001) planes (Fig. S5), where the corresponding peaks are shifted between 6 and 10$^{\circ}$. This might be related to the strain the SRO film experiences in the [001] direction. The matching peak positions indicate that the pseudocubic unit cell of SRO has the same orientation with the STO cubic unit cell. This type of alignment is favorable as it minimizes in-plane strain \cite{jiang1998microstructure}. Similarly, the azimuthal scans for SRO$_{110}$ (Fig. \ref{fig1}B) show coinciding substrate and film peaks indicating that again the alignment of the film unit cell mimics that of the substrate. 

To probe the magnetic anisotropy of the samples, magnetic measurements were done with the magnetic field directed along different crystallographic axes. Magnetization was measured as a function of the field strength with the field along the film normal, at 45$^{\circ}$ to the film normal, and parallel to the film surface. For SRO$_{110}$, two in-plane measurements were conducted as the two primary in-plane axes ([001] and [1$\overline{1}$0]) are inequivalent. Diamagnetic contributions arising from the substrate were subtracted. We conducted these measurements at 10 K, which is far below the Curie temperature (around 150 K as determined from both magnetization-temperature and resistance-temperature measurements (Fig. S6)). 

As seen in Fig. \ref{fig1}C in SRO$_{001}$, the measurements conducted with the field directed along the in-plane direction indicate that this direction is close to a hard axis. The out-of-plane hysteresis loop, on the other hand, has a large saturation magnetization and the largest remanent magnetization while the measurement conducted at 45$^{\circ}$ shows the largest coercive field. This indicates that the easy axis lies between 45$^{\circ}$ and 90$^{\circ}$, meaning that it has an considerable out-of-plane component accompanied by a tilt towards the plane of the film.

For the SRO$_{110}$ samples, the magnetic hysteresis loops were measured with the field aligned with the surface normal, along 45$^{\circ}$ and along the two principal in-plane directions ([001] (blue) and [1$\overline{1}$0] (green)). Fig. \ref{fig1}D shows the different hysteresis loops. It is clear from the results that the out-of-plane direction is the magnetic easy axis, evidenced by the largest remanent magnetization, saturation magnetization and coercivity. The measurement conducted along 45$^{\circ}$ shows a reduced, but still significant, remanence and coercive field. The two in-plane loops have similar shapes with no significant coercive field or remanent magnetization, indicative of hard axes. 

For both films, the high field saturation magnetic moments are different with the field applied along different directions. This is commonly observed in SrRuO$_3$ films and this effect has several possible origins. Ziese \textit{et al.} discussed that the procedure used to subtract the diamagnetic substrate background, done by taking the slope at high fields, does not take into consideration paramagnetic contributions originating from the film itself and hence can lead to different saturation values along different crystallographic directions \cite{Ziese2010}. It is also possible that because the anisotropy field of SRO is higher than the maximum applied field, full saturation is not realized along the magnetic hard axes \cite{Koster2012a}. A third possible explanation is that the presence of defects prevents films from reaching full saturation \cite{Zahradnik2020}.

Ac electrical measurements were performed after Pt deposition and fabrication of Hall bars. In ferromagnetic materials the magnetization orientation contributes to the resistance due to the interactions between the conduction electrons and magnetic moments, giving rise to a collection of magnetoresistance (MR) effects. The MR effects are a convenient tool for understanding various material properties as it is sensitive to for instance the crystalline symmetry, magnetocrystalline anisotropy, defects, domains and domain walls \cite{Ziese2010}. 

In order to analyze and compare these aspects of the different films, we studied MR effects by applying a 7 T magnetic field while rotating the films from in- (0$^{\circ}$) to out-of-plane (90$^{\circ}$). It should be noted that the magnetization does not exactly follow the field orientation but tends towards the easy axes. This allows for further analysis and comparison of the magnetic anisotropies of the SRO/Pt bilayers on the different substrates by studying the first harmonic longitudinal and transverse resistances. The measurements we present here were performed by applying a 1.5 mA current. Further measurements, in which the current amplitude was varied are shown in supplementary section 3, and indicate the onset of discernible second harmonic signals to occur between 0.5 and 1 mA.

The longitudinal resistance ($R$), measured along the current (x) direction, reflects the fact that resistance depends on the angle between magnetization and current. Known as the anisotropic magnetoresistance (AMR), this is described by Eq. \ref{eq1} or Eq. \ref{eq2} depending on whether the field is rotated in the xz or yz plane respectively. $\theta$ indicates the angle between the in-plane direction parallel (xz) or perpendicular to the applied current direction (yz).
\begin{equation}
\label{eq1}
    \Delta R^{xz}_{\omega}=(R^x-R^z)\cos^2\theta
\end{equation}
\begin{equation}
\label{eq2}
    \Delta R^{yz}_{\omega}=(R^y-R^z)\cos^2\theta
\end{equation}
Where $\Delta R$ represents the change in resistance induced by the AMR. In both cases, the sign of the amplitude depends on whether the resistance is higher or lower when the field is applied in-plane with respect to out-of-plane. 
\begin{figure}
    \centering
    \includegraphics[width=0.7\textwidth]{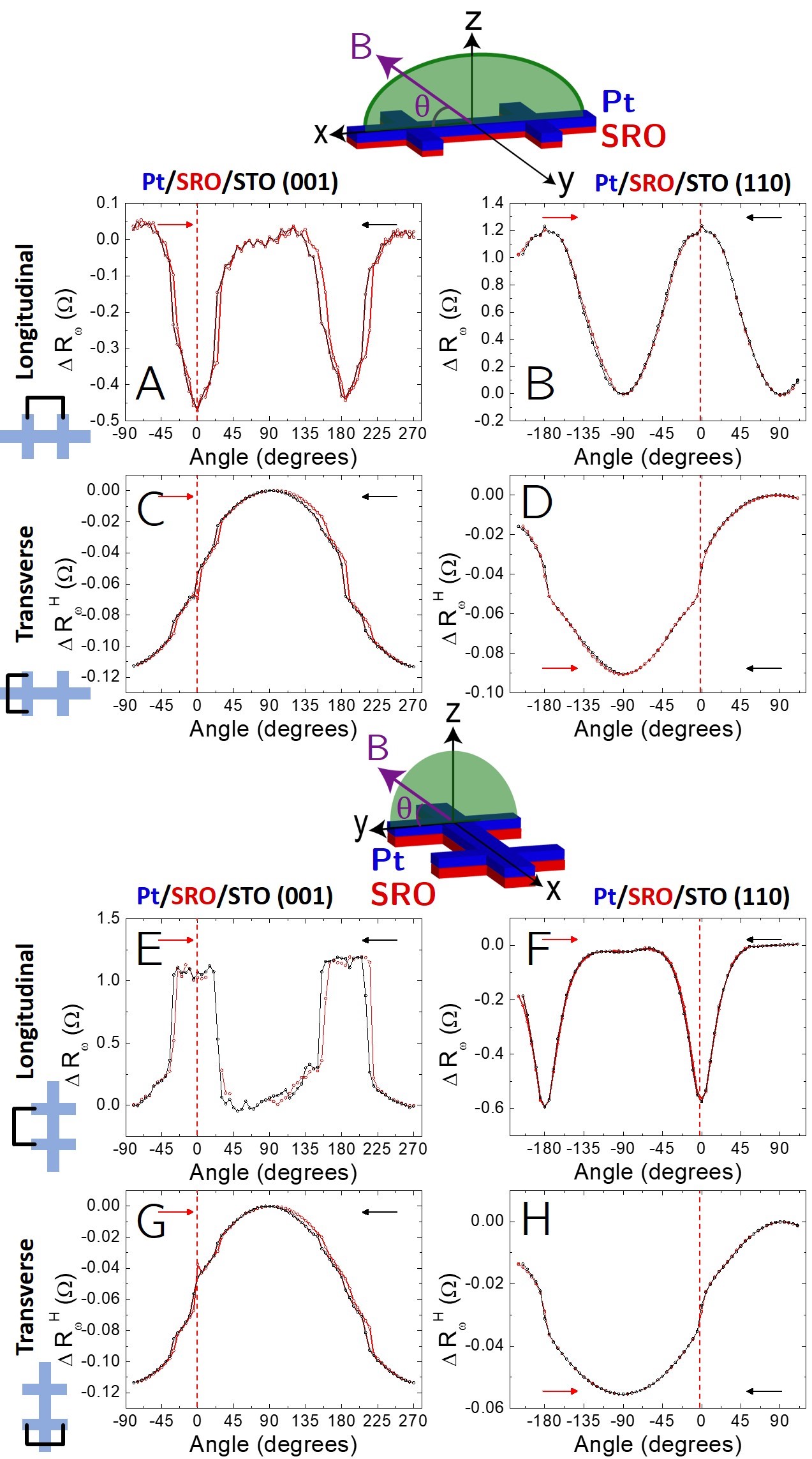}
    \caption{First harmonic angular measurements for SRO$_{001}$ (left) and SRO$_{110}$ (right). Data for xz sweeps are shown in \textbf{(A)} and \textbf{(B} for the longitudinal resistance and in \textbf{(C)} and \textbf{(D)} for the transverse resistance. The yz scans are shown in \textbf{(E)} and \textbf{(F)} for the longitudinal resistance and in \textbf{(G)} and \textbf{(H)} for the transverse resistance. $\Delta R$ indicates the change in resistance with respect to when the field is applied out-of-plane. The angle ($\theta$) of the magnetic field ($B$) is indicated with respect to the in-plane direction. Red and black lines are used to indicate the trace (increasing angles) and retrace (decreasing angles) scans respectively.}
    \label{fig2}
\end{figure}
The field orientation scans in the xz-plane are shown in Fig. \ref{fig2}A (SRO$_{001}$) and B (SRO$_{110}$). For SRO$_{001}$ the longitudinal resistance is minimized when the current and field are aligned and maximized when the field and current are perpendicular. An interesting feature manifesting the anisotropy in the system is the differing shapes of the dips and peaks: the minima are sharp while the maxima are significantly broader. On the other hand, the field orientation scan for SRO$_{110}$ yields behavior that more closely resembles the expected cos$^2\theta$-dependence with similar peak shapes for both the peaks and dips. Moreover, now the maximum longitudinal resistance is observed when the field and current are parallel while the perpendicular alignment yields the minimum resistance. Finally, we observe a small additional sharp feature occurring whenever the current and field point in the same direction. 

Scanning the field orientation in the yz-plane preserves the perpendicular orientation of the field with respect to the current channel. As shown in Fig. \ref{fig2} E (SRO$_{001}$) and F (SRO$_{110}$), for SRO$_{001}$, we observe flat resistance maxima when the magnetization lies in plane and broader dip when the field points out-of-plane. The peaks and dips are separated by sharp transitions that occur when the field is $\pm \sim$30$^{\circ}$ away from one of the in-plane directions. For SRO$_{110}$, on the other hand, we see sharp dips when the field points in-plane and broad peaks centred around the out-of-plane direction.

The transverse resistance ($R_H$)\, measured perpendicular to the current (y) direction, results from the Anomalous Hall Effect (AHE) and as opposed to AMR, is expected to have a $\sin\theta$ dependence. Deviations from the expected smooth behavior are typically a fingerprint of strong magnetocrystalline anisotropy. Similar behavior is observed for both SRO$_{001}$ and SRO$_{110}$ in both xz- and yz-scans (Fig. \ref{fig2} C, D, G and H). In each plot, a maximum transverse resistance is realized whenever the field points along the +z direction (90$^{\circ}$) and minima in resistance when the field points in the -z direction, as expected. However, the transition between one state to the other is relatively sharp resulting in a deviation from the expected smooth sine-dependence. Fig. \ref{fig2} D and H also indicate that for SRO$_{110}$ when the field is aligned in-plane a single sharp transition occurs while for SRO$_{001}$ (Fig. \ref{fig2} C and G) a series of steps are seen spanning a larger range of angles around 0$^{\circ}$ and 180$^{\circ}$. 

In addition to first harmonic signals, we studied second harmonic responses. As well as linear effects, the voltage response also contains higher order terms due to the nonlinear current-induced response of the system. The spin-orbit torques, for instance, scale linearly with the current and therefore the SOT-related voltage signal is expected to scale quadratically with current. The effect of current-induced spin-orbit torques on the magnetization direction can therefore be probed with second harmonic component of the voltage \cite{Garello2013,avci2015}. It should be noted that also thermal effects scale linearly with current and therefore manifest in the second harmonic voltage signal. Overall, studying the variation of the second harmonic signals as a function of the field orientation allows identifying current-induced modulation of the magnetization, whether of thermal or non-thermal origin. 

\begin{figure}
    \centering
    \includegraphics[width=0.7\textwidth]{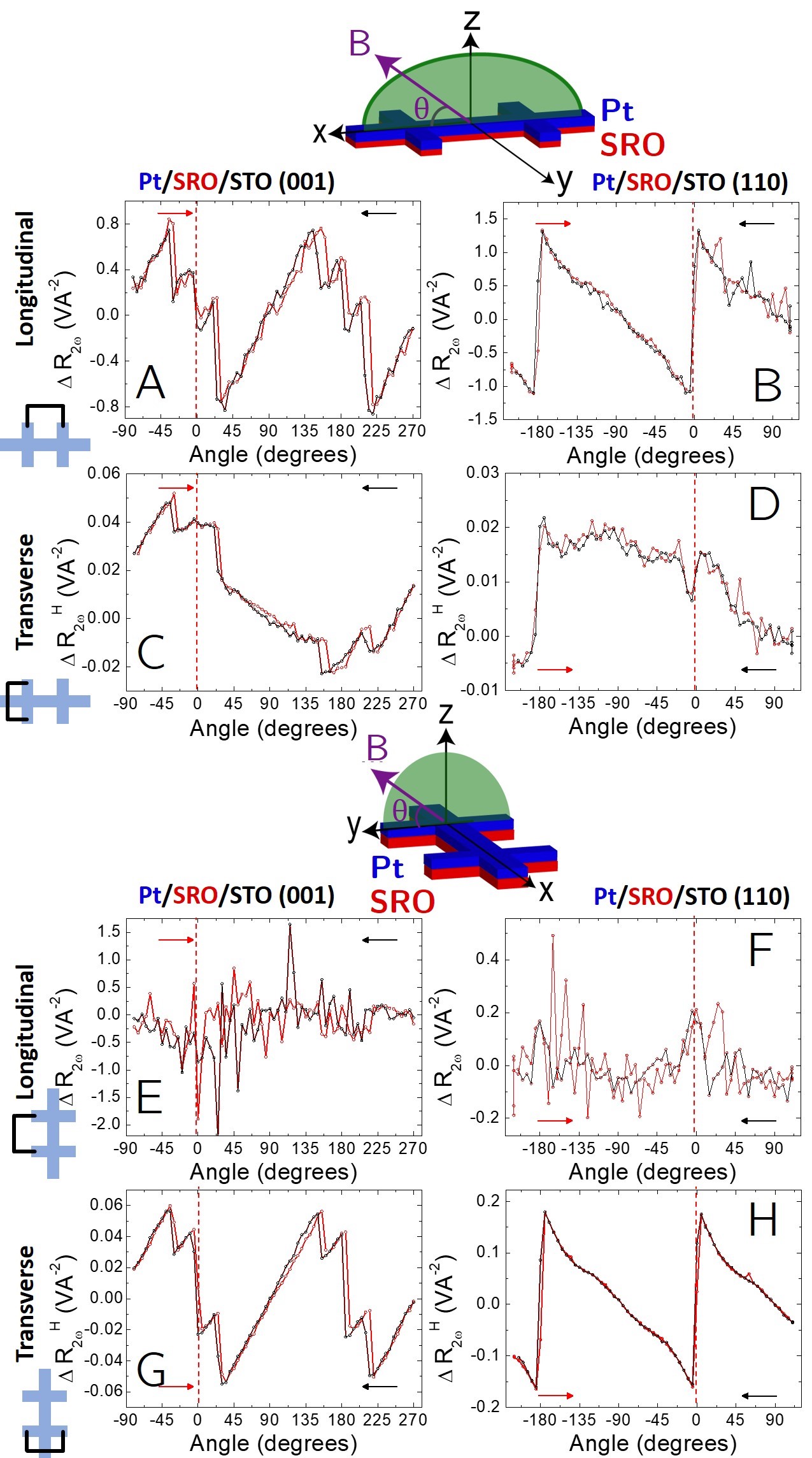}
    \caption{Second harmonic angular measurements for SRO$_{001}$ (left) and SRO$_{110}$ (right). Data for xz sweeps are shown in \textbf{(A)} and \textbf{(B)} for the longitudinal resistance and in \textbf{(C)} and \textbf{(D)} for the transverse resistance. The yz scans are shown in \textbf{(E)} and \textbf{(F)} for the longitudinal resistance and in \textbf{(G)} and \textbf{(H)} for the transverse resistance. The angle ($\theta$) of the magnetic field ($B$) is indicated with respect to the in-plane direction. $\Delta R$ indicates the change in resistance with respect to when the field is applied out-of-plane. Red and black lines are used to indicate the trace (increasing angles) and retrace (decreasing angles) scans respectively.}
    \label{fig3}
\end{figure}

The longitudinal second harmonic data when rotating the field in the xz plane are shown for SRO$_{001}$ and SRO$_{110}$ in Fig. \ref{fig3} A and B respectively. Both graphs reveal strong and clear signals with a 180$^{\circ}$ periodicity. In both cases, the angular dependencies comprise of a (series of) vertical jump(s) between the signal maximum and minimum as well as a sloped line between a minimum and maximum. For SRO$_{110}$, a sharp jump spanning only a narrow range of angles occurs whenever current and field are parallel. For SRO$_{001}$, on the other hand, a series of three jumps and cusps appears between the peak and dip. The transition from peak to dip is still centred around 0$^{\circ}$, but spans a wider range of angles of around 60$^{\circ}$. On the other hand, no clear signals are observed for either heterostructure when the magnetic field is rotated in the yz plane as seen in Fig. \ref{fig3} E (SRO$_{001}$) and F (SRO$_{110}$). This can be due to a too low signal-to-noise ratio or the absence of angular dependence. 

The transverse signals for xz plane rotations are significantly different. Both signals appear to have a 360$^{\circ}$ periodicity, but their shapes differ. For SRO$_{001}$ there is a peak in the signal at 0$^{\circ}$ and a minimum at 180$^{\circ}$. Jumps and cusps are present around a relatively wide range of angles surrounding the situation when the field points in-plane. For SRO$_{110}$, a rapid change in the signal is seen at 180$^{\circ}$ and a small jump-like feature appears when the angle is 0$^{\circ}$, given that this feature appears consistently between trace and retrace this is not believed to be an artefact. 

The transverse signals for rotations in the yz plane yield clear angular dependencies for both samples (Fig. \ref{fig3} G and H). The signal to noise ratio is remarkably high for both. Qualitatively, the angular variations are similar for the longitudinal signals in the xz scans. The periodicity of the signals is 180$^{\circ}$ and the graphs show (a series of) vertical jump(s) between global dips and peaks. For SRO$_{110}$, a single sharp jump is observed when the field is in-plane, while for SRO$_{001}$, a series of jumps spanning a wider range of angles is seen.

\section{Discussion}
For SRO grown on STO (001) the XRD results showed that for both the substrate and film (in pseudocubic notation) the out-of-plane axes is [001], and the in-plane axes of the film also align with the equivalent axes of the STO. Similar conclusions can be drawn for SRO grown on STO (110), where again the film axes are aligned with the equivalent substrate axes. As a result, the [110]$_{pc}$ SRO crystal axes, i.e. the expected magnetic easy axes, for SRO$_{001}$ and SRO$_{110}$ lie in different spatial directions. This is confirmed by the magnetic measurements shown in Fig. \ref{fig1} that indicate the easy axis of SRO$_{110}$ to lie along the film normal, while that of SRO$_{001}$ is tilted away from the film normal. The combined results of XRD and magnetization studies are summarized in Fig. \ref{fig1}  E and F, indicating the unit cell orientations and magnetic easy axes of the different samples. 
The longitudinal resistance is predominately attributed to AMR arising from conducting through the metallic SRO. It is typically observed that in heavy metal/ferromagnetic metal systems contributions from AMR are significantly larger than those arising from spin Hall magnetoresistance \cite{karwacki2020}.
AMR is typically composed of two components (crystalline and non-crystalline) that have different microscopic origins. The non-crystalline component originates from the influence of the relative angle between the magnetization and current on the transport scattering matrix elements. The crystalline component, on the other hand, stems from spin-orbit coupling and arises because the rotating magnetic field induces changes in the equilibrium relativistic electronic structure \cite{mcguire1975anisotropic,rushforth2007anisotropic}. 

The non-crystalline AMR dominates in the xz scans shown in Fig. \ref{fig2} A and C. For SRO$_{001}$ we observe that the resistance is lower when the current and field are parallel than when they are perpendicular. This is commonly referred to as negative AMR. While it is opposite to what is found in the most common ferromagnets, it has been observed in SRO and several other systems \cite{Herranz2004a}. Given that the spin-polarization of SRO is -9.5$\%$ \cite{Worledge2000,Herranz2003a}, the dominance of minority spin transport could be the cause of the negative AMR \cite{Ziese2000,Takata2017}. 

Deviations from the expected cos$^2\theta$ shape are likely the result of anisotropy. In particular, the sharpness of the dips that occur when the field is in-plane indicates this is close to a magnetic hard axis. On the other hand, the broad and flat peaks centred around the out-of-plane directions indicate the vicinity of a magnetic easy axis. The flattened maxima might further indicate that there are multiple easy axes that lie close to, and on either side of the out-of-plane directions with small tilts. Rather than following the field, the magnetization may jump between these axes. 

For SRO$_{110}$, a more conventional positive AMR is seen, reflecting the fact that the shape of atomic orbitals depends on the direction of magnetization. In particular, the scattering cross-section is larger when the current and field are parallel giving rise to maximum resistance while the cross-section is smallest when the field and current are orthogonal.

During the rotational measurements conducted in the yz-plane, in which the angle between the field and current is constant, the AMR is expected to be more sensitive to the crystalline component. For SRO$_{001}$, flat resistance maxima appear when the magnetization lies in-plane, and sharp transitions when the field is $\pm \sim$30$^{\circ}$ away from one of the in-plane directions. The rapid jump in resistance suggests a magnetic easy axis lying perpendicular to the angle, namely oriented approximately 30$^{\circ}$ away from the film normal. The broad dip spans a similar range of angles as the peaks in Fig. \ref{fig2}A, again corroborating the existence of easy axes tilted away from the film normal.

The sharp dips in SRO$_{110}$ when the field is in-plane reflect that the in-plane direction is magnetically hard. The easy axis in this case is expected to lie perfectly out-of-plane. The sharpness of the dips in this scan is in contrast to the broad, but flattened peaks, and cannot be explained considering only low order anisotropy terms. Instead, higher order crystalline anisotropy terms should be taken into consideration to fully explain the complex AMR. 

The transverse resistance measurements for both samples yield similar results, but with some important differences. The expected sine periodicity is observed where the highest resistance is found whenever the field points up (+z) and the lowest resistance when the field points down (-z). This indicates that for both substrates the AHE resistance is positive, consistent with the out-of-plane field scans shown in Fig. S13. The shape deviates from that of a perfect sine as a result of the transitions that occur when the field approaches the in-plane direction. The single sharp transition seen at 0$^{\circ}$ and 180$^{\circ}$ for SRO$_{110}$ suggests in-plane to be a hard magnetic plane and it to be more favorable for the magnetization to lie out-of-plane. For SRO$_{001}$ the transition region spans a greater range of angles and multiple jumps and cusps are seen to occur in this region. This is consistent with a more complex anisotropy and multiple easy axes with a small tilt away from the film normal. 

The longitudinal second harmonic data for xz scans (Fig. \ref{fig3} A and B) yields strong signals with similar features for both samples. One of the significant differences is that for SRO$_{110}$, a single sharp jump occurs when the current and field are parallel while for SRO$_{001}$ this transition comprises of three jumps and cusps spanning a wider range of angles of about 60$^{\circ}$ around the in-plane configuration. This is consistent with the first harmonic transverse results in which a series of (SRO$_{001}$) and a single (SRO$_{110}$) jump(s) are seen for the different samples. As mentioned previously, this is likely a signal of the differences in anisotropy, in particular with SRO$_{110}$ possessing well-defined uniaxial anisotropy and SRO$_{001}$ having multiple easy axes. No longitudinal second harmonic signals were seen when instead the magnetic field was swept in the yz plane, where the field is always perpendicular to the current. The symmetry of the longitudinal xz signal and lack of yz signal is consistent with the presence of anti-damping like SOT, with a small contribution likely due to thermal gradient \cite{avci2015,han2020}.

The transverse signals for xz plane rotations have a 360$^{\circ}$ periodicity. The yz plane rotations yield clear angular dependencies with a period of 180$^{\circ}$. These signals could originate from SOT, but there can also be a contribution from a vertical thermal gradient. The high signal-to-noise ratio of the second harmonic signals likely originates from spin-orbit torque in both Pt/SRO/STO samples. Section 2 in the supplementary information utilizes the four-direction method proposed by Park \textit{et al.} \cite{Park2019} to eliminate thermoelectric contributions from the harmonic Hall signals. The line shapes are not significantly altered by this method, indicating SOT to be the dominant origin of the observed signals. 

\subsection{Spintronic Memristor}
\begin{figure}
    \centering
    \includegraphics[width=\textwidth]{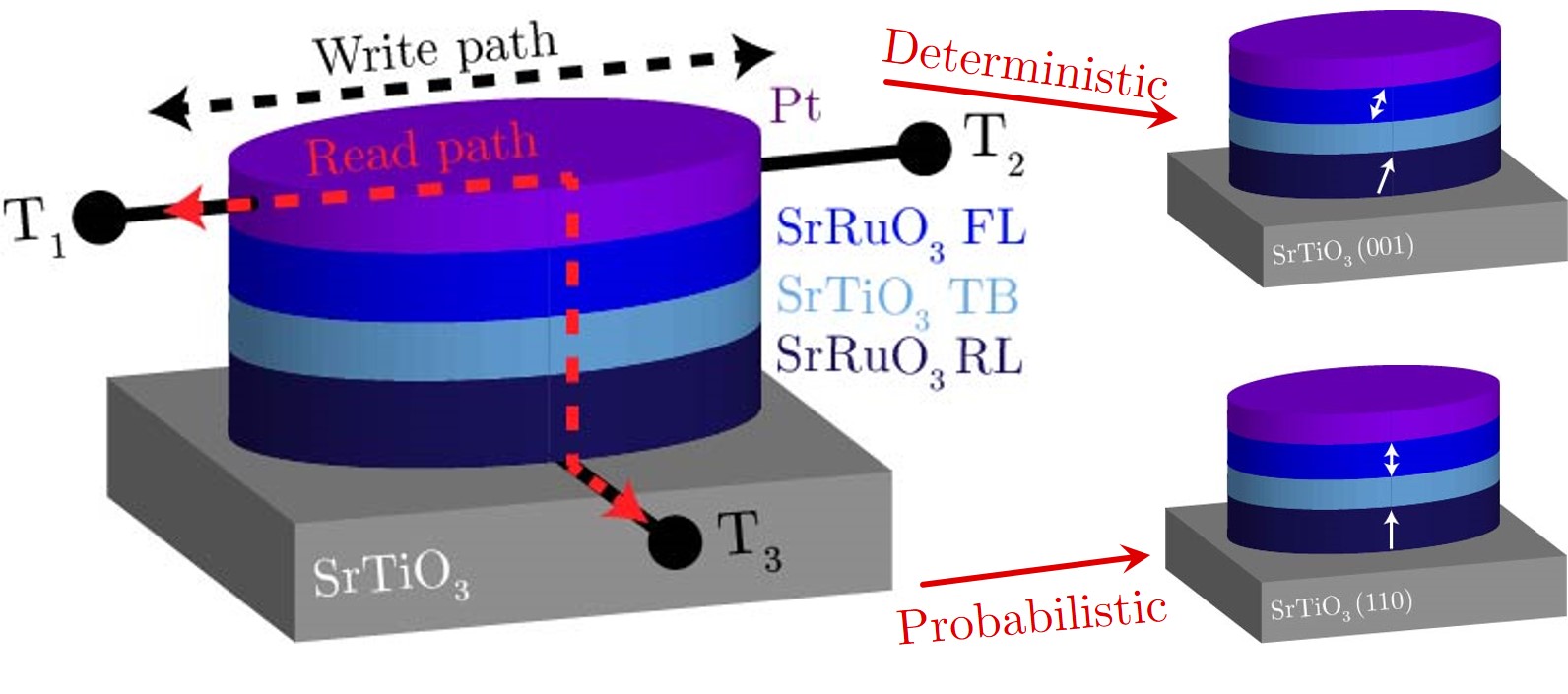}
    \caption{ Schematic of the proposed device structure for neuromorphic spintronic memristors. The device consists of an STO substrate/SRO reference layer (RL)/STO tunnel barrier (TB)/SRO free layer (FL)/Pt. The write path is between terminals $T_1$ and $T_2$ and the read path between $T_1$ and $T_3$. The right side of the figure indicates how the choice of substrate dictates whether the device will show deterministic or probabilistic behavior.}
    \label{fig4}
\end{figure}
Finally, we discuss potential applications of these SRO/STO systems in the field of neuromorphic computing. When the easy axis is along the film normal, as is found for SRO$_{110}$, an in-plane magnetic field in the direction of the applied current is needed for the magnetization switching to be deterministic. In SRO$_{001}$, where the easy axis is tilted, on the other hand, the symmetry is broken with respect to the in-plane SOT; this makes it possible to achieve field-free deterministic switching in this case. When current is applied through the Pt layer, a spin accumulation will be induced through the spin Hall effect. This, in turn, exerts a spin-transfer torque on the magnetic moments of SRO. With a sufficiently high current density this torque can force the magnetization to align with the spin accumulation direction \cite{Eason2013,You2015,Liu2019}.

Memristors are circuit elements that can show several resistive states controlled by the amount of electric charge. These devices are deemed particularly useful for realising neuromorphic computing, because their behavior is reminiscent of that of synapses, and they have also been used to emulate neurons. The magnetization state in these SRO films is continuously tunable and can be used as memristors when included as the free layer in a magnetic tunnel junction (MTJ). Here the state of magnetization can be read by tunnel magnetoresistance. A proposed device structure for a three-terminal spintronic synapse consists of two transistors and one MTJ device (2T1MTJ) \cite{Sakimura2007,Fukami2017}. A schematic of the MTJ device is shown on the left hand side of Fig. \ref{fig4}. The device consists of an STO substrate/SRO reference layer (RL)/STO tunnel barrier (TB)/SRO free layer (FL)/Pt. Three terminals are required as the read and write paths are electronically separated. The writing path is between terminals $T_1$ and $T_2$, and the read path uses terminals $T_1$ and $T_3$. The isolation of the read and write paths requires two transistors which may be disadvantageous, it enables fast and reliable operation.

A useful benefit of utilising spintronic devices is that the possibility of magnetization control through both magnetic field as well as current allows both individual devices to be addressed (by current) in addition to all devices to simultaneously be set to the same state (using magnetic field). Based on this, we propose a three-terminal synaptic MTJ device in which different non-volatile states can be written by applying an in-plane current. The memory can then be read out by a small out-of-plane current. Introducing the third terminal has several practical advantages. Firstly, it is more energy efficient to separate the read and write paths as writing through the tunnel barrier would require large amounts of current to be passed though an insulating layer, resulting in heat generation. Secondly, by passing large currents through the tunnel barrier it is possible that over time damage occurs, leading to limited endurance. 

The right side of Fig. \ref{fig4} indicates how the choice of substrate dictates whether the device will show deterministic or probabilistic behavior. Using STO (001) substrates we showed that we can realize an easy axis which is slightly tilted away from the film normal, hence breaking the symmetry and allowing field-free deterministic switching in a relatively simple device structure without additional layers. This device could serve as a functional spintronic synapse. In neural networks, synapses typically play the role of representing a weight, which can be viewed as the equivalence of a conductance regulating the flow of information between the two neurons it connects. For a network to function properly, these weights should be non-volatile and able to take on various levels within a range. It should also be possible to controllably change the conductance with an external stimulus. In this case, the weight could be controlled by direct current (dc) pulses - the direction in which the pulse is sent will determine whether it potentiates or depresses the weight of the synapse. 

Similarly when STO (110) is used and the easy axis is oriented fully out-of-plane, the probabilistic nature of switching can be utilized to get different functionalities out of the device. 
Probabilistic behavior is becoming increasingly important as stochastic systems are being used for unconventional computing \cite{lv2017,hayakawa2021}. Stochastic MTJs have, for example, been proposed for integer factorization as an alternative to qubits (\cite{borders2019}) and image segmentation \cite{liyanagedera2017}. Sourcing a write current will result in probabilistic switching of the free magnetization, where the probability is determined by the current magnitude \cite{Srinivasan2016,Shim2017}. This could act, for example, as a non-linear activation function or a spiking spin-neuron in a network. Hence, by controlling the magnetic anisotropy of the system, the same materials can be used to potentially realize both neuronal and synaptic devices.

In order to use such a device utilizing SRO as a free layer, an oxide tunnel barrier and a second ferromagnetic PMA layer as the reference layer is needed. Using SRO for the latter as well would be advantageous as again no additional layers are needed to induce an out-of-plane easy axis. Such MTJ stacks have been demonstrated by Herranz \textit{et al.} consisting of uncoupled SRO/STO/SRO stacks on STO substrates in which both SRO layers exhibit PMA \cite{Herranz2003a}.
The benefit of this approach lies in the simplicity of the device design compared to conventional PMA devices whose operation relies on a multitude of layers.

\section{Conclusions}
In this work, we demonstrate the ability to control magnetic anisotropy of SrRuO$_3$ ferromagnetic layers by the choice of substrate. The tailored anisotropy can potentially allow both probabilistic and deterministic current-induced magnetization switching when a Pt layer is added. In light of neuromorphic applications, this gives the possibility to realize two different spintronics memristive devices where the anisotropy controls the device functionality. With perfect perpendicular anisotropy, a device with probabilistic switching could be made which can serve the role of an artificial neuron. By tuning the easy axis to have a slight tilt, the switching can be deterministic and provide synaptic functionality. 

\subsection*{Conflict of Interest Statement}
The authors declare that the research was conducted in the absence of any commercial or financial relationships that could be construed as a potential conflict of interest.

\subsection*{Author Contributions}
A.G. and M.L. fabricated the samples carried out device processing and performed characterization and electrical measurements. The data were interpreted and the paper was written by A.G., M.L. and T.B..

\subsection*{Acknowledgments}
The authors would like to thank all members of the Spintronics of Functional Materials group at the University of Groningen, in particular P. Zhang for her guidance in PLD growth of SRO films and A. A. Burema for ion beam etching.
Device fabrication was realized using NanoLab NL facilities. 
The authors acknowledge technical support from J. Baas, J. G. Holstein, H. H. de Vries, A. Joshua, T. Schouten, and H. Adema. 
A.S.G. is supported by the the CogniGron Centre, University of Groningen.



\printbibliography

\clearpage
\renewcommand{\figurename}{Figure S}
\renewcommand\thefigure{\arabic{figure}}
\setcounter{figure}{0} 

\section*{Supplementary Data}

\section{Structural and Magnetic Data}

\begin{figure}[H]
    \centering
    \includegraphics[width=0.5\textwidth]{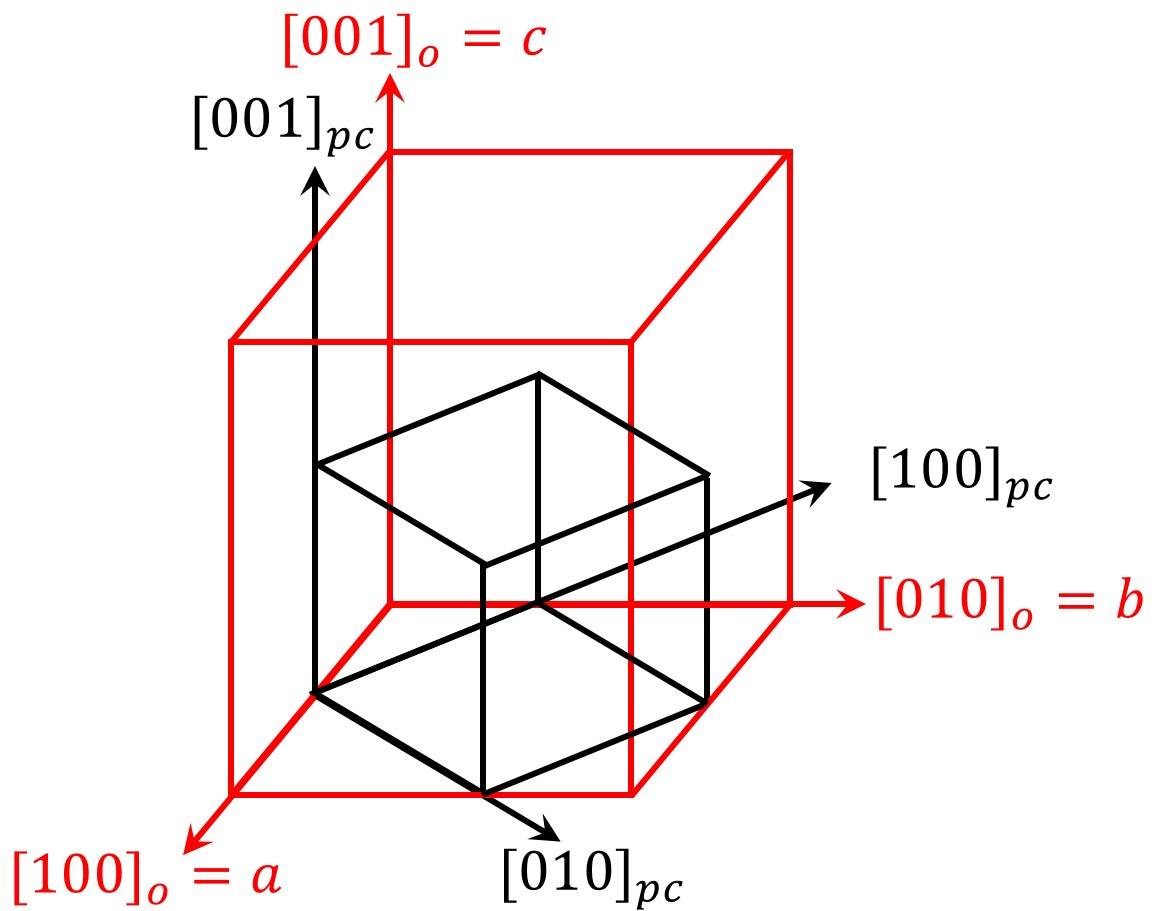}
    \caption{Schematic indicating the relative orientations of the unit cells and important crystallographic axes in the orthorhombic (red) and pseudocubic (black) unit cell notation of SrRuO$_3$.}
    \label{figs1}
\end{figure}

\begin{figure}[H]
    \centering
    \includegraphics[width=\textwidth]{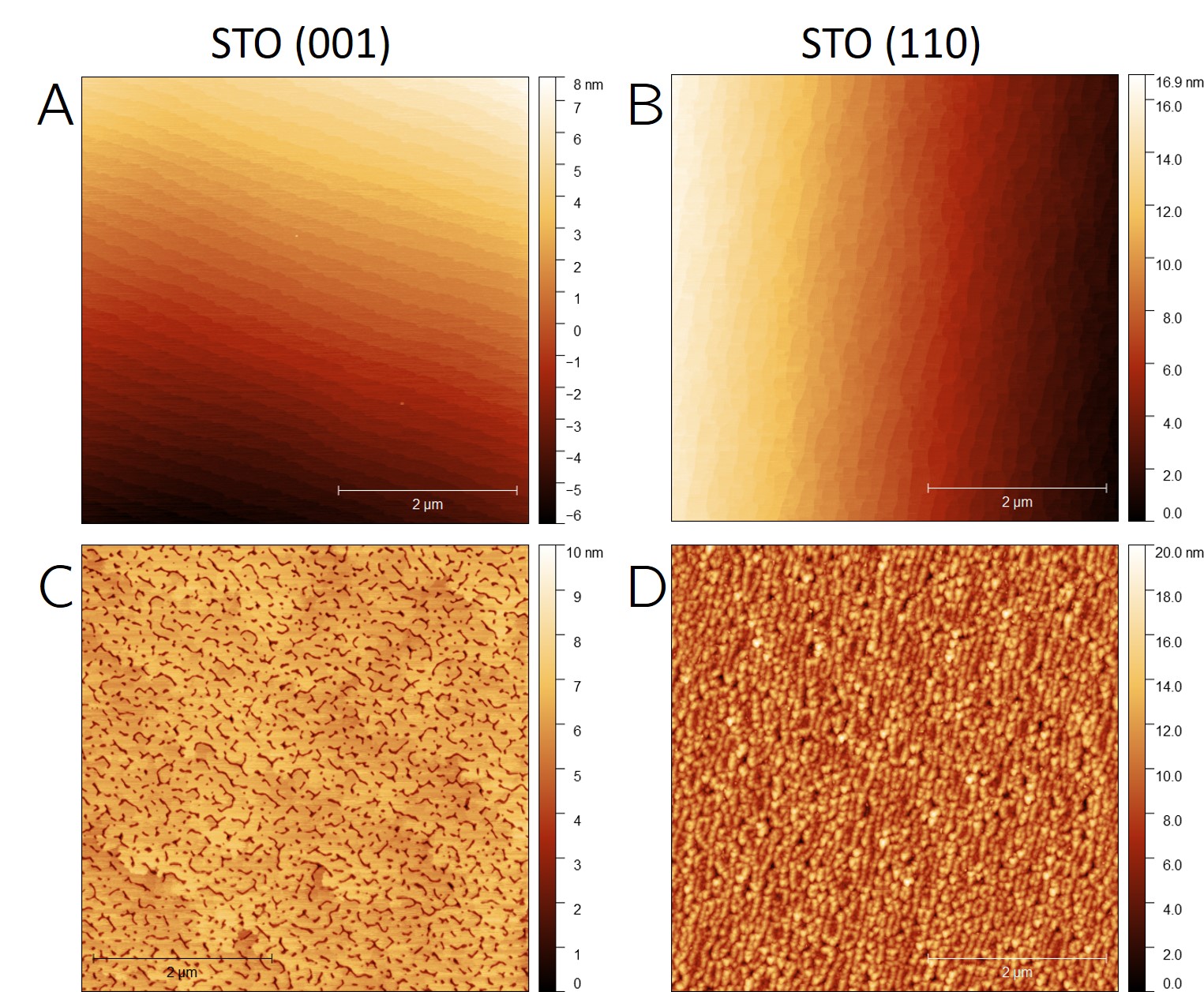}
    \caption{Atomic force microscopy (AFM) images of the SrTiO$_3$ (001) substrate (\textbf{A}) after chemical termination and annealing and (\textbf{C}) after SrRuO$_3$ deposition. AFM images of STO (110) (\textbf{B}) after termination and annealing and (\textbf{D}) after SRO deposition.}
    \label{figs2}
\end{figure}
\begin{figure}[H]
    \centering
    \includegraphics[width=\textwidth]{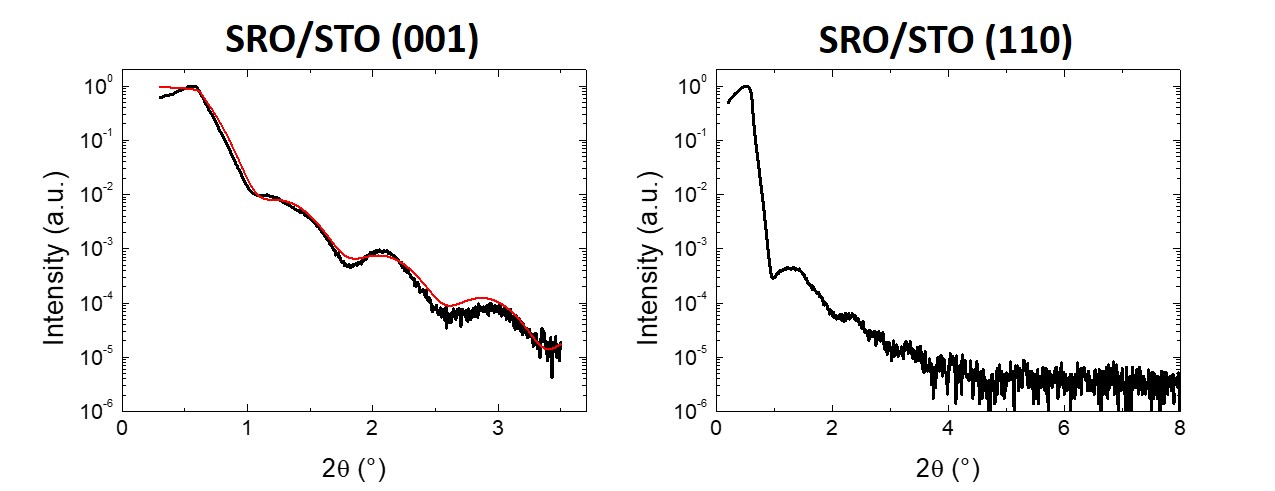}
    \caption{X-ray reflectivity data for \textbf{(A)} SRO/STO(001) and \textbf{(B)} SRO/STO (110). For \textbf{(A)} the fit using Parratt's formulism is shown in red.}
    \label{figs3}
\end{figure}
\begin{figure}[H]
    \centering
    \includegraphics[width=\textwidth]{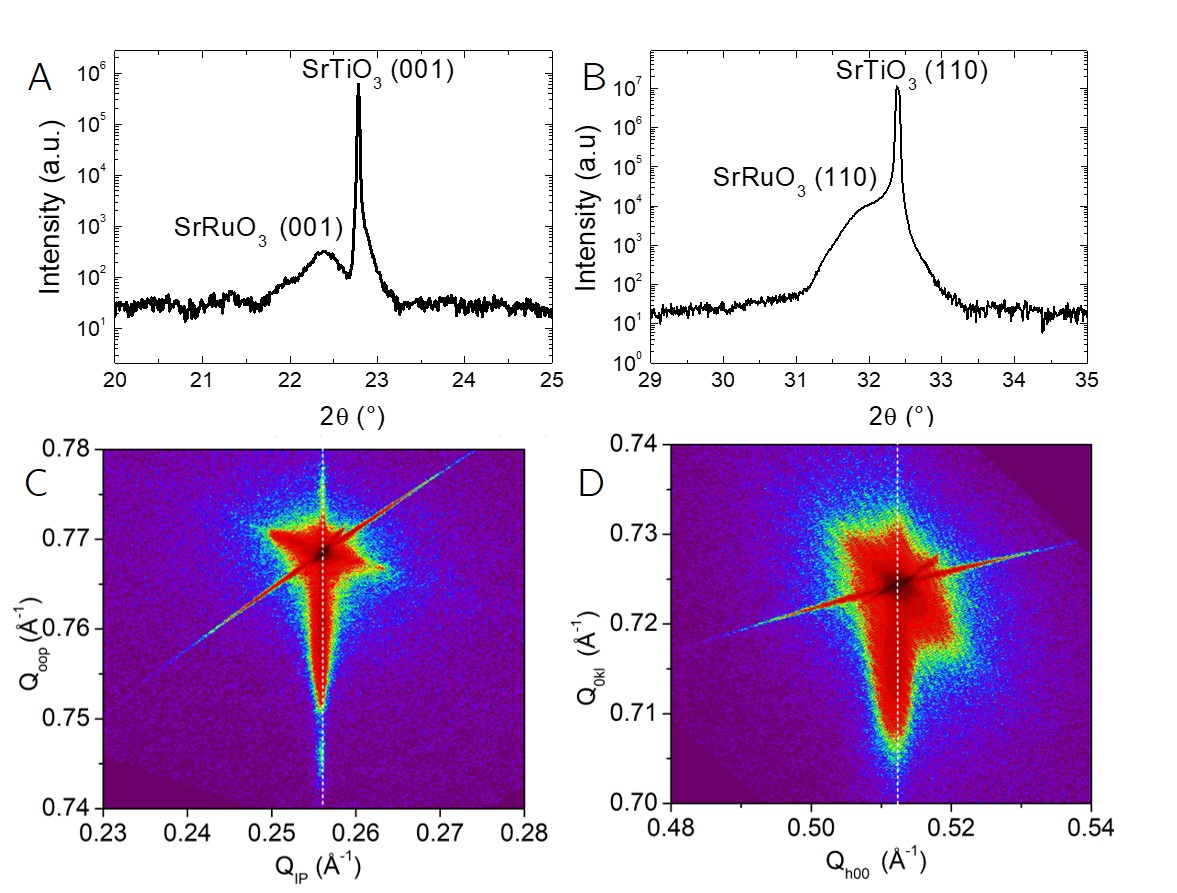}
    \caption{2$\theta$-$\theta$-scans around the (001) SRO and STO peaks for SRO/STO (001) \textbf{(A)} and around the (110) SRO and STO peaks for SRO$_{110}$ \textbf{(B)}. Reciprocal space maps \textbf{(C)} for SRO$_{110}$ around the (103) peak and \textbf{(D)} for SRO$_{110}$ around the (222) peak.}
    \label{figs4}
\end{figure}
\begin{figure}[H]
    \centering
    \includegraphics[width=\textwidth]{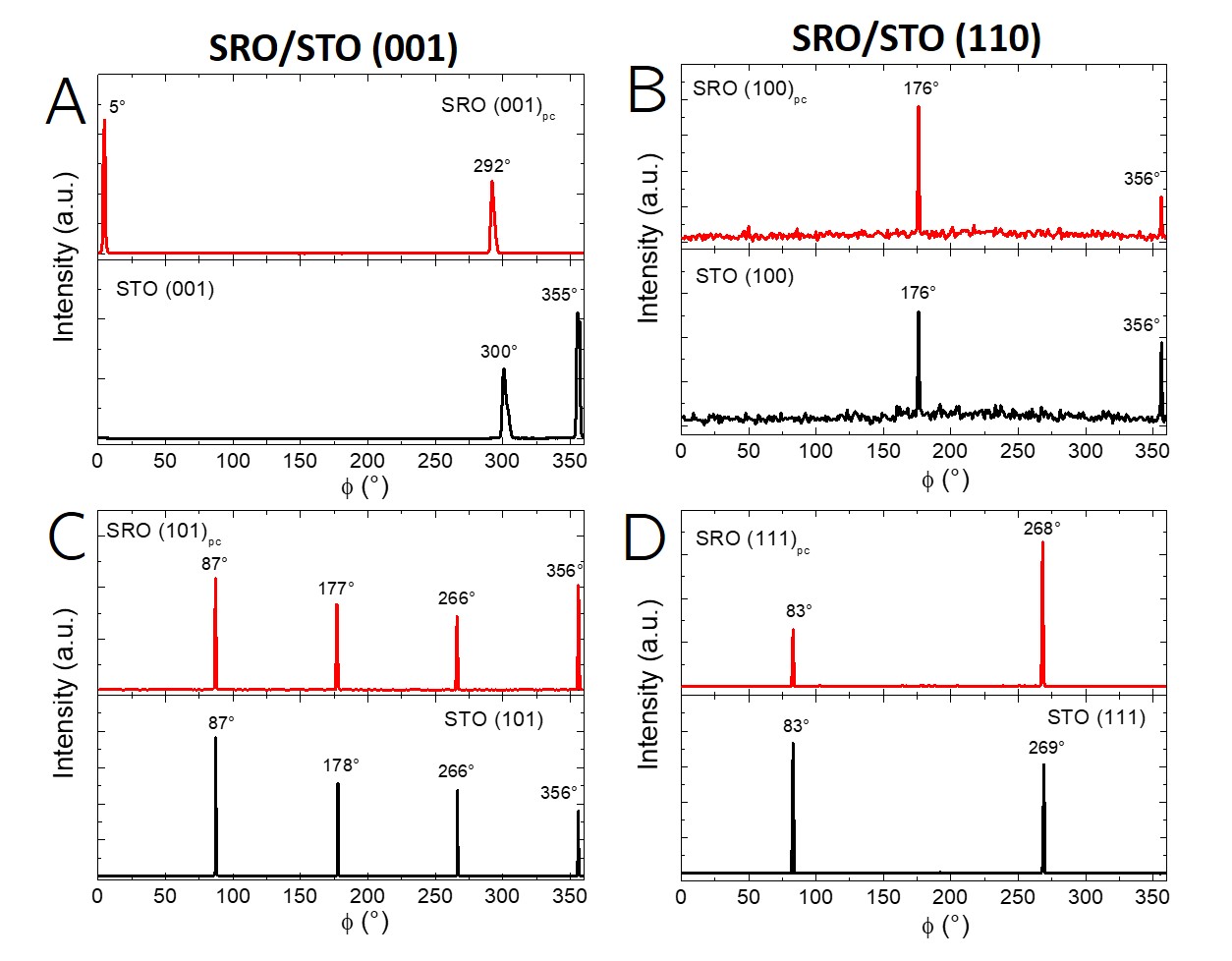}
    \caption{Azimuthal scans for SRO/STO (001) (left) and SRO/STO (110) (right) around different lattice points. (\textbf{A}) (001), (\textbf{B}) 101, (\textbf{C}) 100 and (\textbf{D}) 111.}
    \label{figs5}
\end{figure}
\begin{figure}[H]
    \centering
    \includegraphics[width=0.9\textwidth]{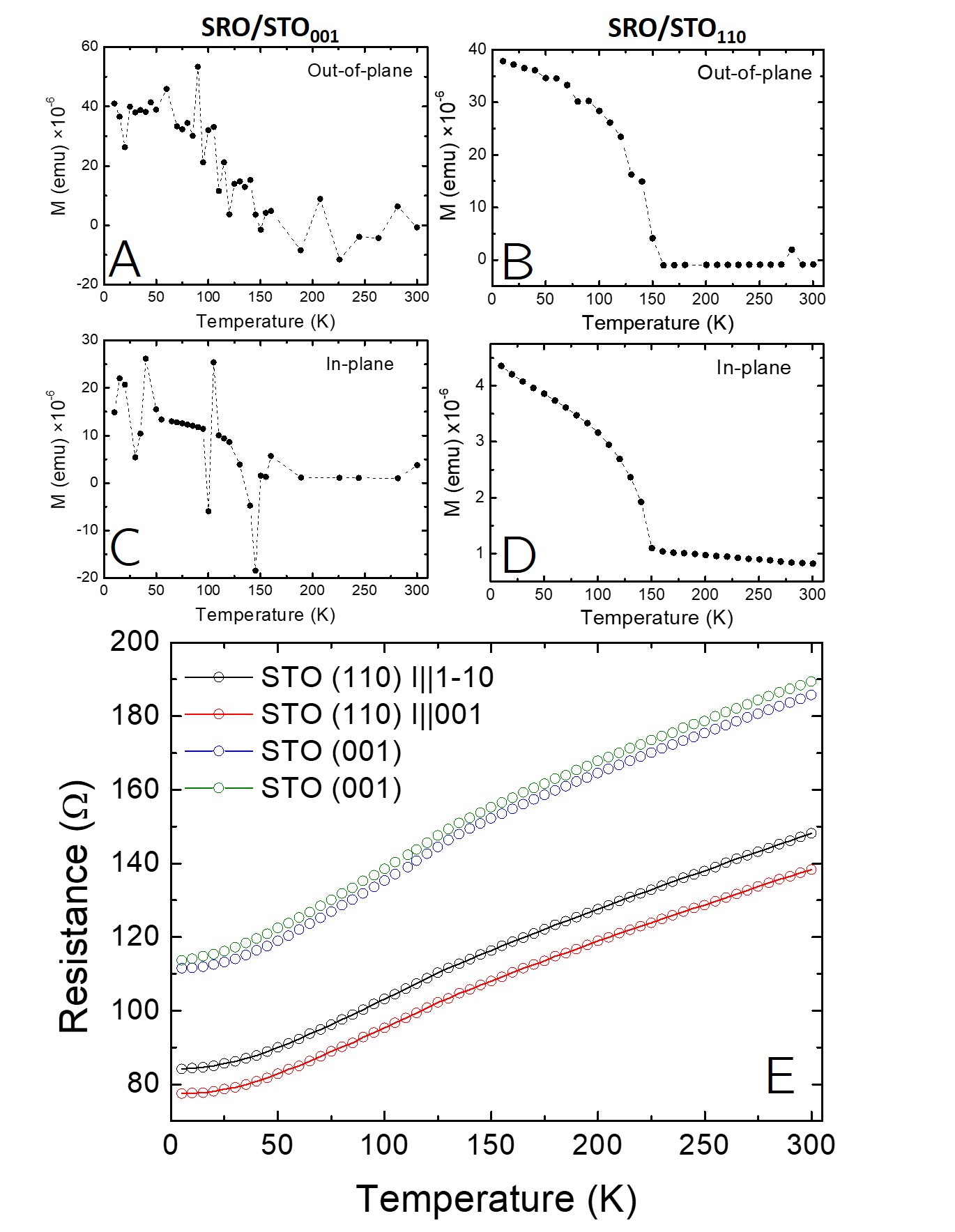}
    \caption{Magnetization-temperature measurements after high field cooling for SRO/STO (001) (left) and SRO (110) (right). The magnetic field was aligned along the film normal ((\textbf{A}) and (\textbf{B})) and in the plane of the film ((\textbf{C}) and (\textbf{D})). (\textbf{E}) resistance versus temperature measurements for differently oriented Hall bars used to perform magnetoresistance measurements.}
    \label{figs6}
\end{figure}
Fig. S5 shows the resistance versus temperature for perpendicularly oriented Hall bars on the different substrates. In each case, a slope change can be  the ferromagnetic transition. No clear differences are seen on the STO (001) substrates, which is to be expected as the two primary in-plane axes are crystalographically identical. For the STO (110) substrate, a small deviation is seen between the two principal directions. This is likely related to the fact that here the current flows along different crystallographic directions.

\section{Elimination of Thermal Effects}
The harmonic Hall measurements include various thermoelectric contributions, Park \textit{et al.} [46] proposed a method to eliminate major thermoelectric signals. This method works for sufficiently large fields where no hysteresis is seen in the second harmonic signal, as is the case for the data presented here. A current injected into a Hall bar generates a temperature gradient giving rise to thermoelectric contributions. First harmonic signals indicate the equilibrium magnetization direction and second harmonic signal show SOT-induced magnetization tilting

The first harmonic Hall voltage is sensitive to contributions from the Seebeck ($V_S$), Nernst ($V_N$), Ettingshaussen ($V_E$), Righi-Leduc effects ($V_R$), as well as an Ohmic offset ($V_O$). These effects can be removed by taking their symmetries wit respect to current and magnetic field into account. This requires four measurements to be taken with different current and field polarities, giving rise to four $V_{1\omega}^{IB}$ terms where superscripts $I$ and $B$ refer to the current and field polarities respectively.
\begin{equation}
\label{hall}
    \frac{V_{1\omega}^{++}-V_{1\omega}^{+-}-V_{1\omega}^{-+}+V_{1\omega}^{--}}{4}=V_{AH}+V_H+V_E
\end{equation}
$V_E$ is usually negligible, allowing for the Hall and anomalous Hall contributions to be isolated. 

Similarly for the second harmonic signals:
Damping-like SOT geometry (B$\parallel$x):
\begin{equation}
\label{xz}
    \frac{V_{2\omega}^{++}-V_{2\omega}^{+-}+V_{2\omega}^{-+}-V_{2\omega}^{--}}{4}=V_{DLT}+V_R
\end{equation}
$V_R$ is a second order effect and is consequently negligible.
Field-like SOT geometry (B$\parallel$y):
\begin{equation}
\label{yz}
    \frac{V_{2\omega}^{++}+V_{2\omega}^{+-}+V_{2\omega}^{-+}+V_{2\omega}^{--}}{4}=V_{FLT}+V_S+V_O
\end{equation}
$V_S$ and $V_O$ contribute a constant offset that can readily be subtracted to isolate the effect of field-like torque.

Figs S7-S10 A show the first harmonic Hall voltages with the 4 different measurement geometries. Using Eq. \ref{hall}, the thermoelectric contributions were eliminated and the results of this are shown in Figs S7-S10 B.
Figs S7 and S9 C shows the second harmonic Hall voltages measured using the 4 geometries for xz rotations. The thermal contributions were removed using Eq. \ref{xz}, giving rise to the graphs in Figs S7 and S9 D. Figs S8 and S10 C shows the second harmonic Hall voltages measured using the 4 geometries for yz rotations. Here, Eq. \ref{yz} was used to to remove the thermoelectric contributions, resulting in the graphs shown in Figs S8 and S10 D.
This analysis indicated that the line shapes of the first and second harmonic Hall signals are not significantly altered by subtracting the thermoelectric artefacts.
\begin{figure}[H]
    \centering
    \includegraphics[width=\textwidth]{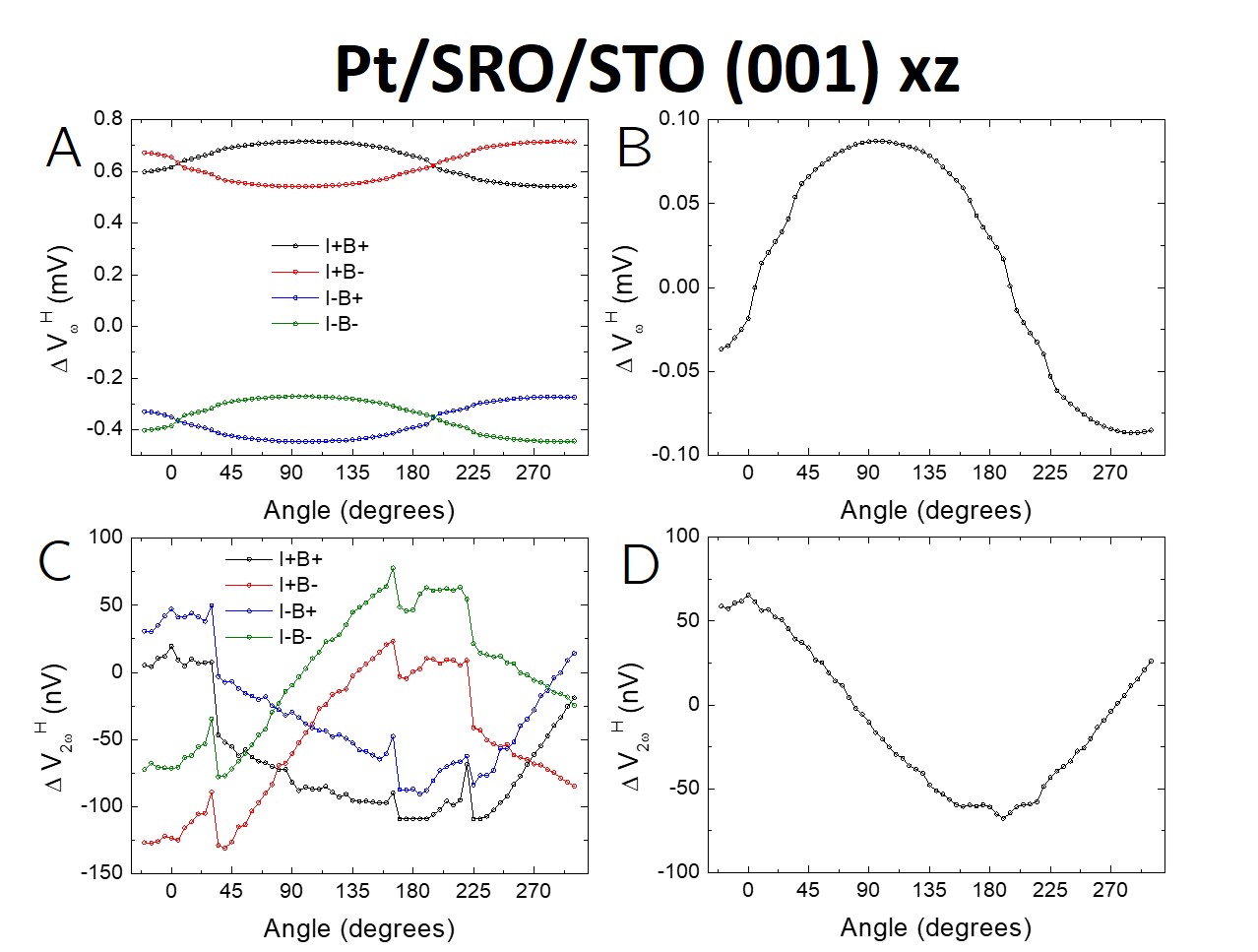}
    \caption{First (\textbf{A}) and second (\textbf{C}) harmonic Hall voltage angle scans taken with a field of 7 T and 1.5 mA for xz scans on Pt/SRO/STO (001). (\textbf{B}) and (\textbf{D}) show the data after eliminating thermal effects according to Eqs. \ref{hall} and \ref{xz} respectively.}
    \label{figs7}
\end{figure}
\begin{figure}[H]
    \centering
    \includegraphics[width=\textwidth]{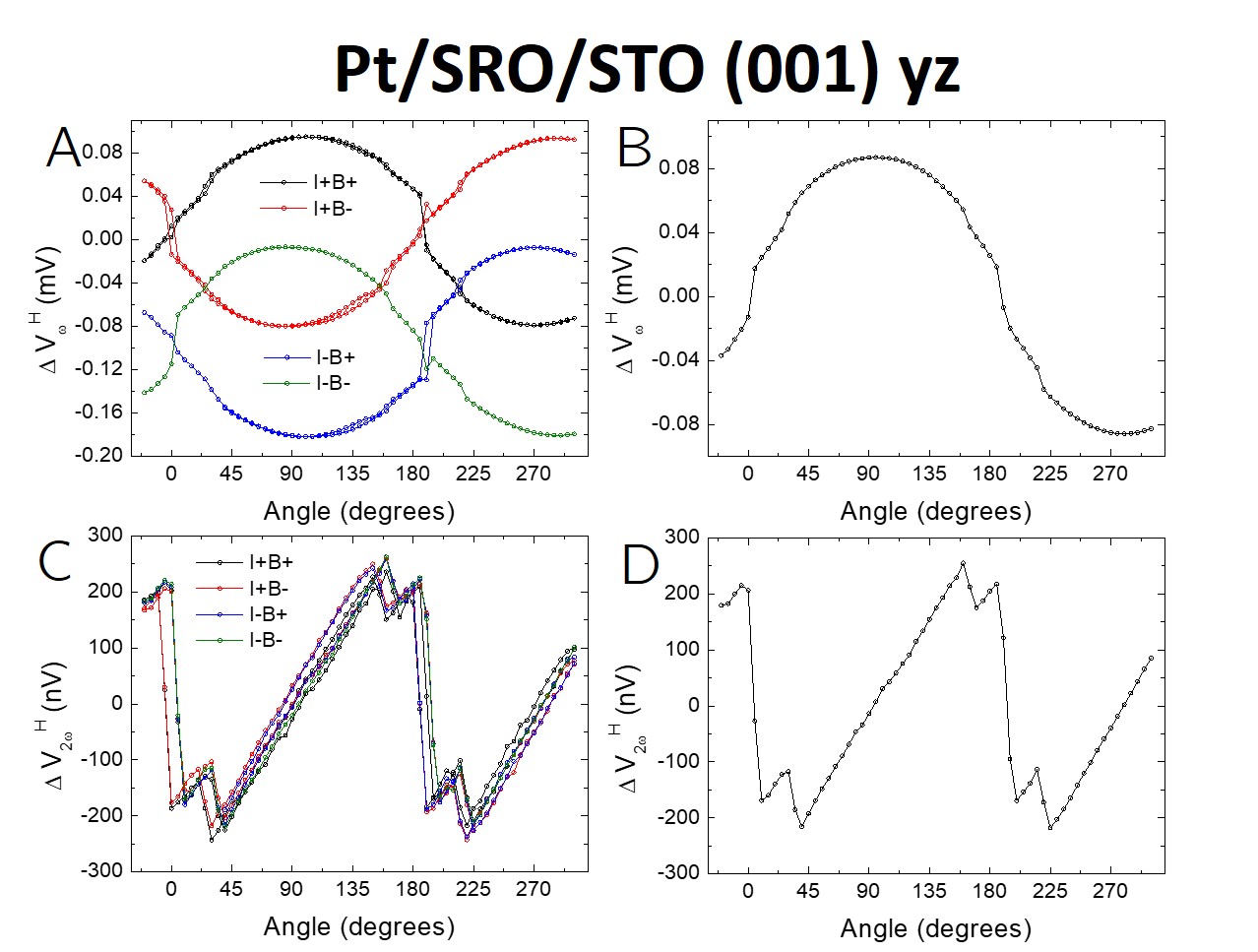}
    \caption{First (\textbf{A}) and second (\textbf{C}) harmonic Hall voltage angle scans taken with a field of 7 T and 1.5 mA for yz scans on Pt/SRO/STO (001). (\textbf{B}) and (\textbf{D}) show the data after eliminating thermal effects according to Eqs. \ref{hall} and \ref{yz} respectively.}
    \label{figs8}
\end{figure}
\begin{figure}[H]
    \centering
    \includegraphics[width=\textwidth]{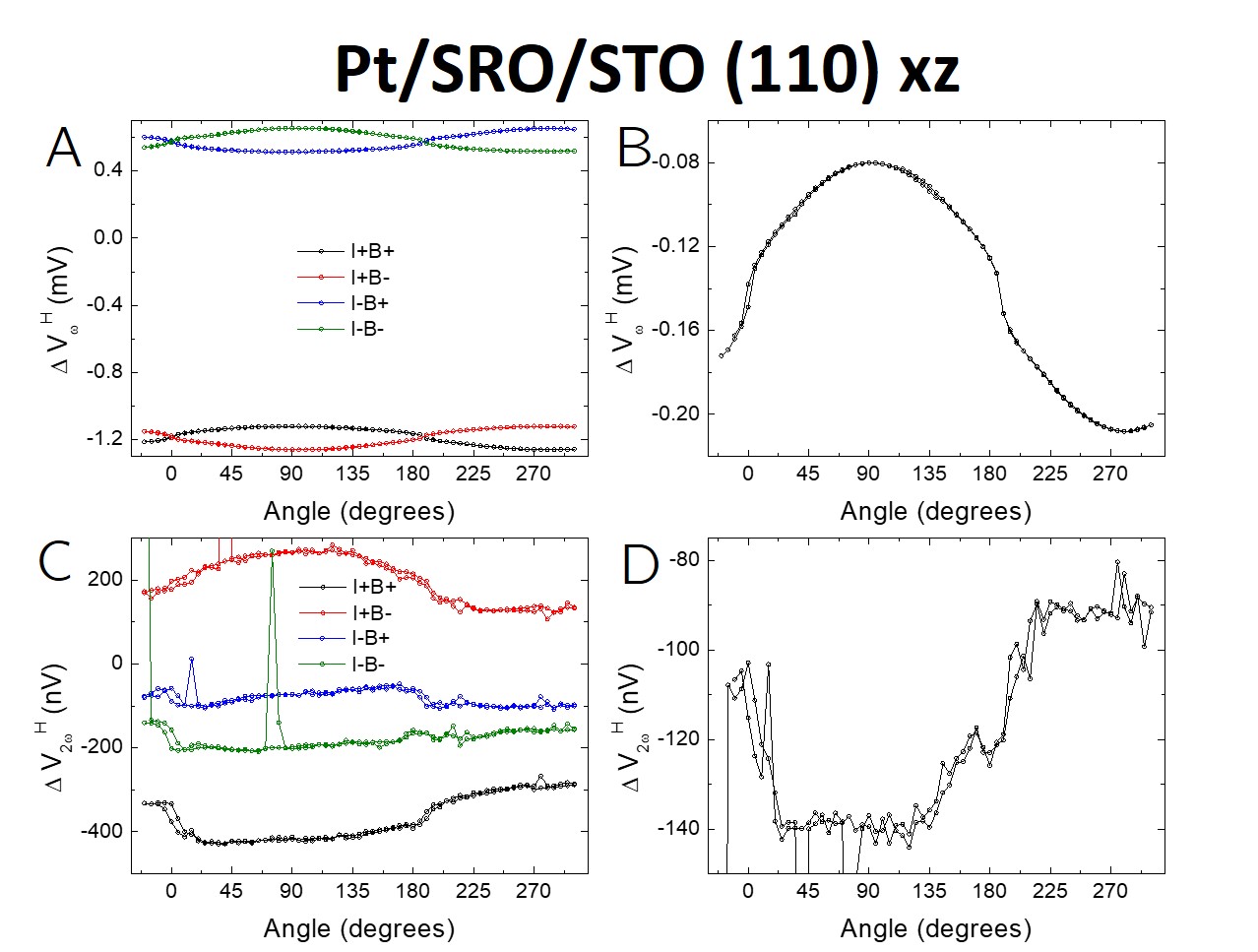}
    \caption{First (\textbf{A}) and second (\textbf{C}) harmonic Hall voltage angle scans taken with a field of 7 T and 1.5 mA for xz scans on Pt/SRO/STO (110). (\textbf{B}) and (\textbf{D}) show the data after eliminating thermal effects according to Eqs. \ref{hall} and \ref{xz} respectively.}
    \label{figs9}
\end{figure}
\begin{figure}[H]
    \centering
    \includegraphics[width=\textwidth]{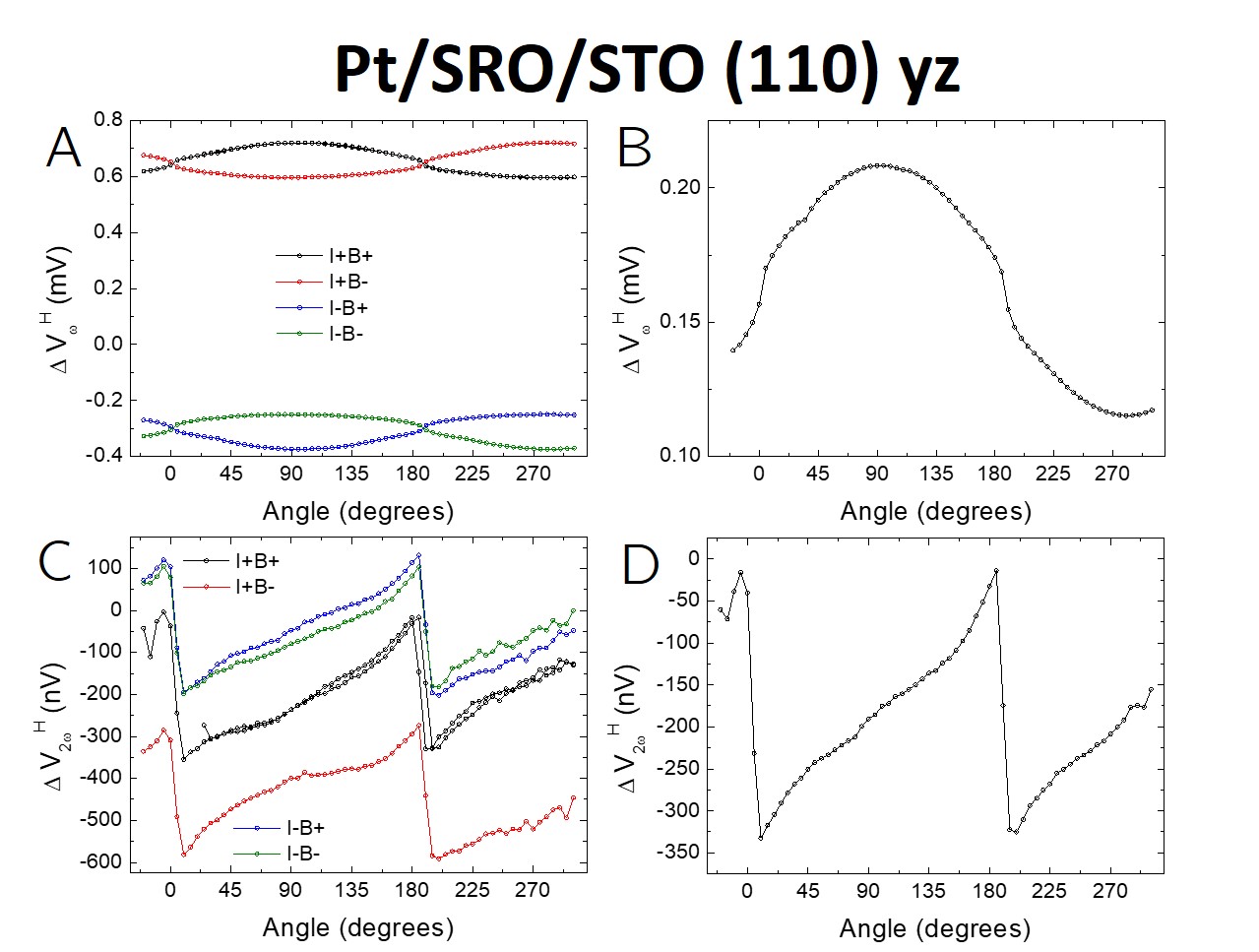}
    \caption{First (\textbf{A}) and second (\textbf{C}) harmonic Hall voltage angle scans taken with a field of 7 T and 1.5 mA for xz scans on Pt/SRO/STO (110). (\textbf{B}) and (\textbf{D}) show the data after eliminating thermal effects according to Eqs. \ref{hall} and \ref{yz} respectively.}
    \label{figs10}
\end{figure}
\section{Current Dependence of Second Harmonic Signals}

\begin{figure}[H]
    \centering
    \includegraphics[width=\textwidth]{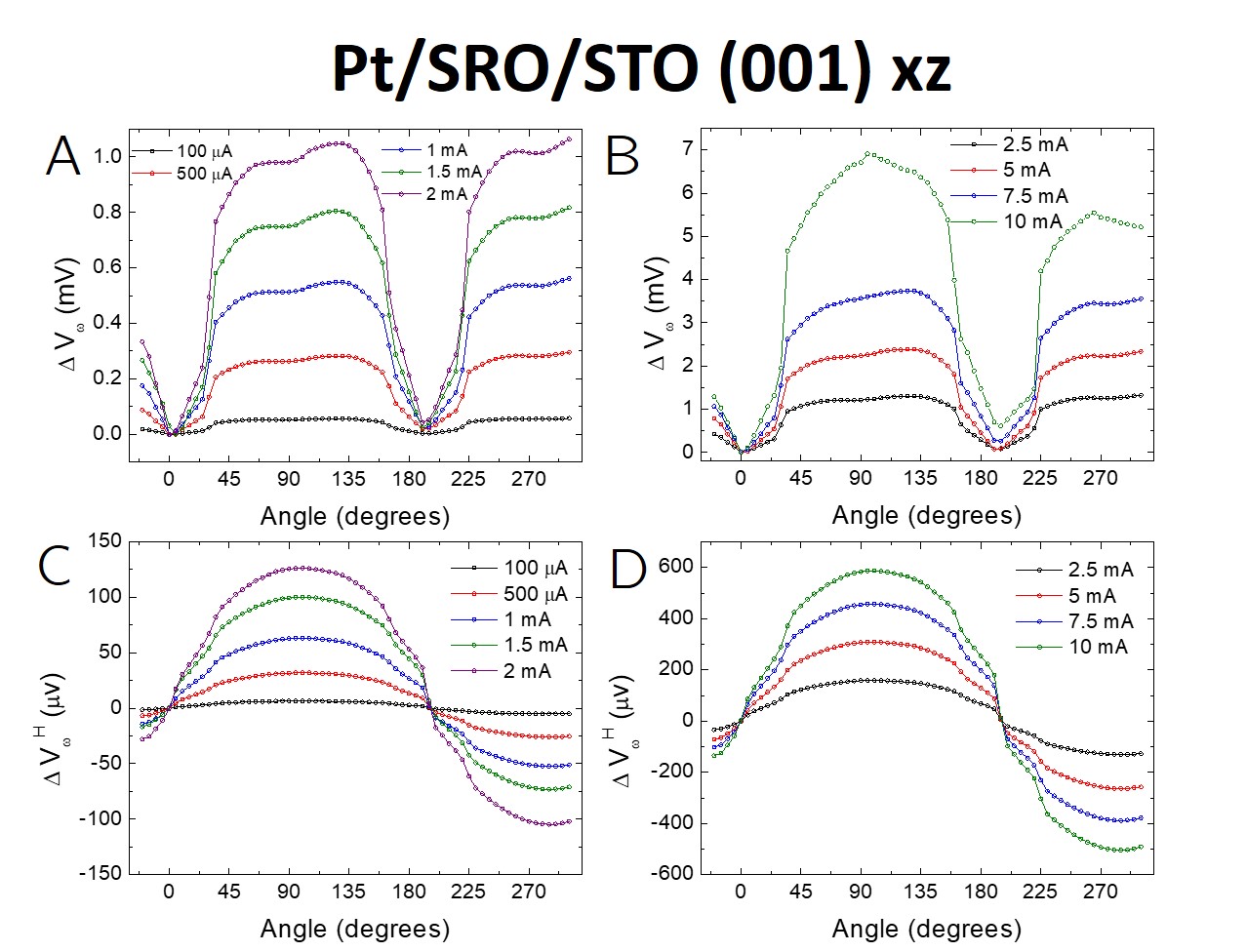}
    \caption{First harmonic signals for Pt/SRO/STO (001) xz rotations for (\textbf{A}) and (\textbf{B}) longitudinal measurements and (\textbf{C}) and (\textbf{D}) transverse measurements with varying currents from 100 $\mu$m to 10 mA.}
    \label{figs11}
\end{figure}

\begin{figure}[H]
    \centering
    \includegraphics[width=\textwidth]{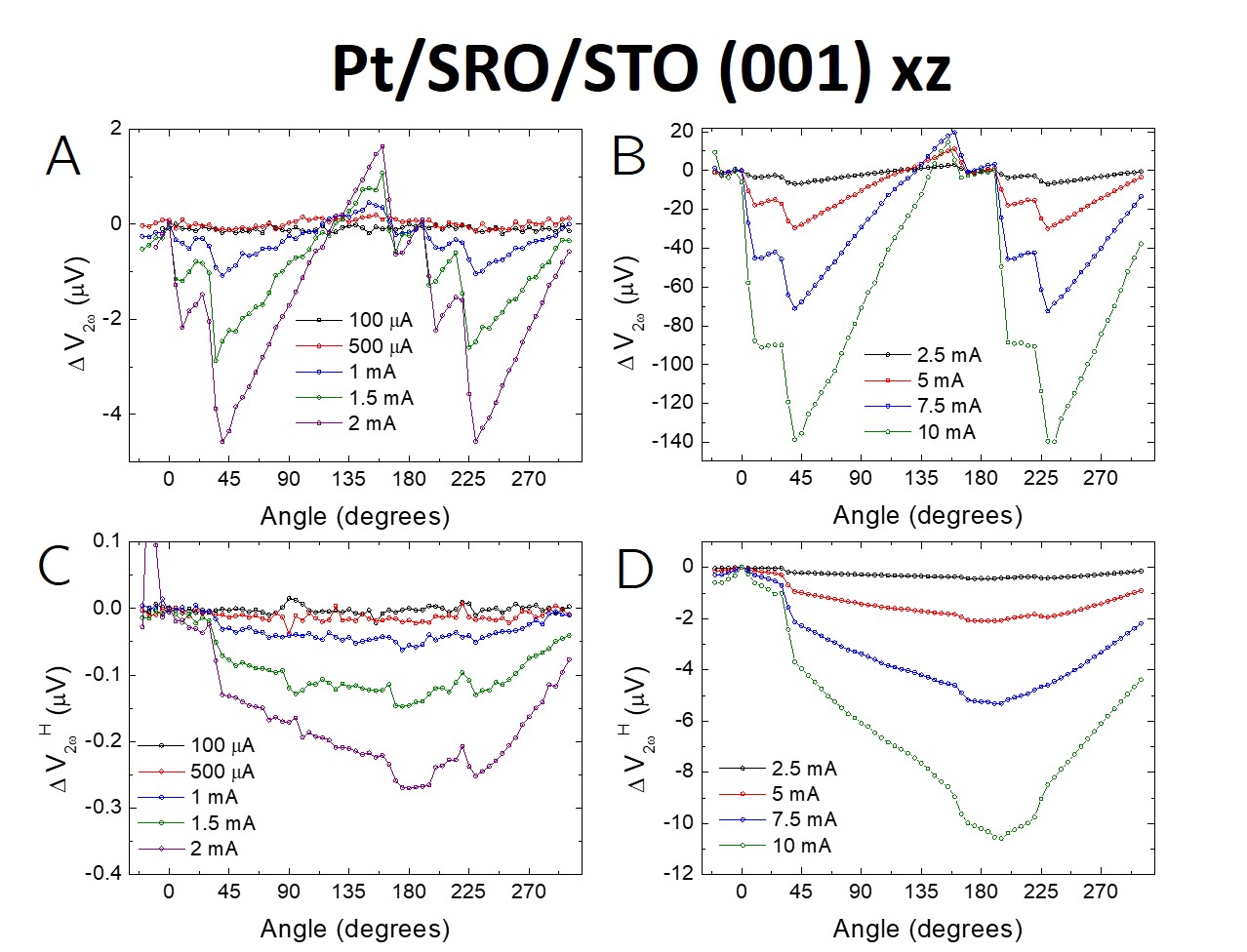}
    \caption{Second harmonic signals for Pt/SRO/STO (001) xz rotations for (\textbf{A}) and (\textbf{B}) longitudinal measurements and (\textbf{C}) and (\textbf{D}) transverse measurements with varying currents from 100 $\mu$m to 10 mA.}
    \label{figs12}
\end{figure}

\begin{figure}[H]
    \centering
    \includegraphics[width=\textwidth]{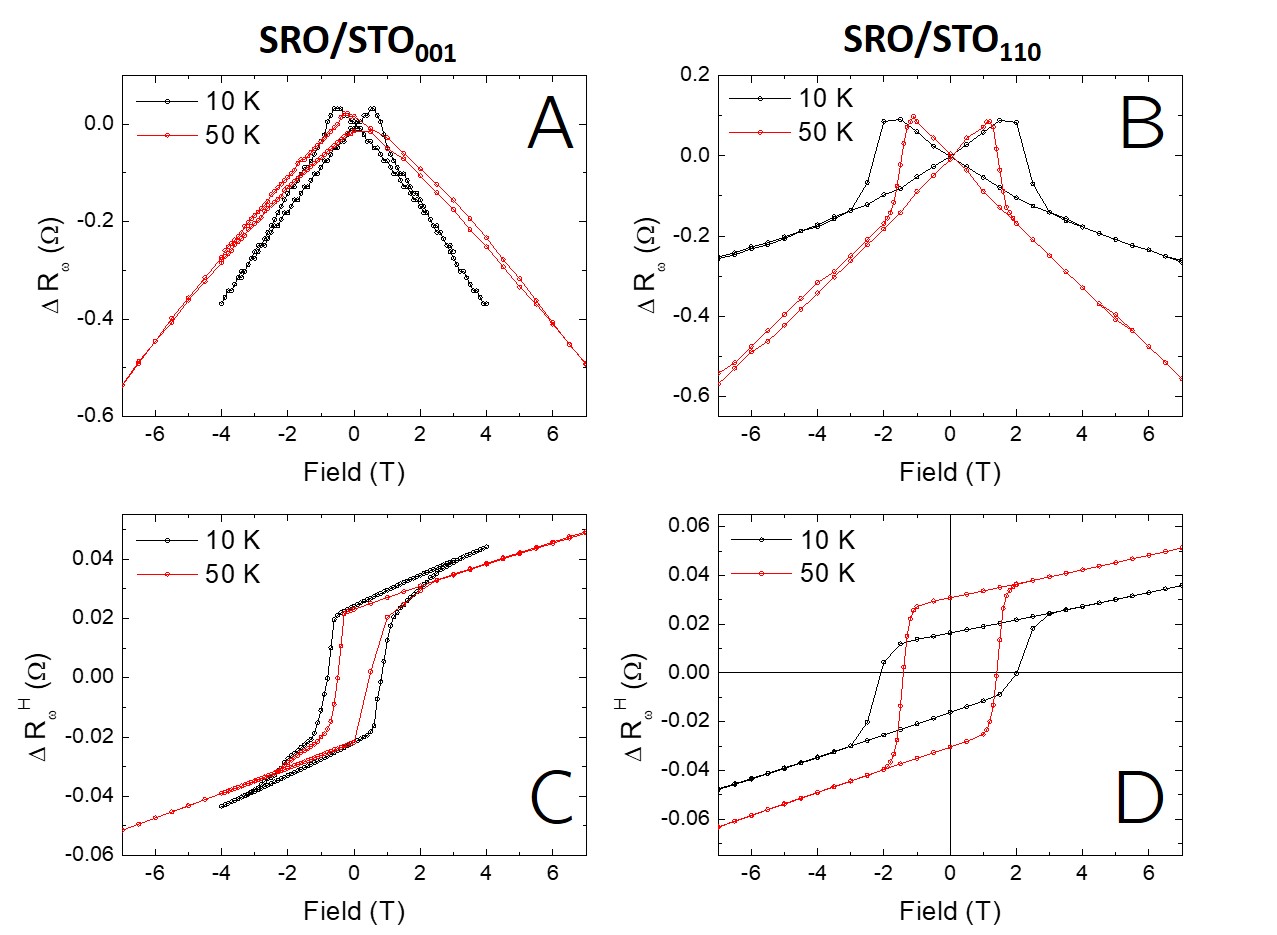}
    \caption{First harmonic resistance versus field measurements with the field along the film normal taken at 10 K (black) and 50 K (red) for SRO/STO (001) (left) and SRO/STO (110) (right). (\textbf{A}) and (\textbf{B}) show the longitudinal resistance and (\textbf{C}) and (\textbf{D}) show the transverse signals.}
    \label{figs13}
\end{figure}

\end{document}